\documentclass{jfm}

\usepackage{graphicx}
\usepackage{epstopdf,epsfig}
\usepackage{newtxtext}
\usepackage{newtxmath}
\usepackage{natbib}
\usepackage{hyperref}
\hypersetup{
    colorlinks = true,
    urlcolor   = blue,
    citecolor  = black,
}

\usepackage{upgreek}
\usepackage{color}
\usepackage[dvipsnames]{xcolor}
\usepackage{tikz}
\usepackage{afterpage}

\usepackage{pdflscape}
\usepackage{rotating}
\usepackage[export]{adjustbox}

\usetikzlibrary{shapes}
\usetikzlibrary{plotmarks}

\graphicspath{{./Figures/}}

\renewcommand{\theenumi}{\alph{enumi}}

\definecolor{LightGray}{rgb}{0.75,0.75,0.75}
\definecolor{c1}{rgb}{0.11,0.62,0.47}
\definecolor{c2}{rgb}{0.85,0.37,0.01}
\definecolor{c3}{rgb}{0.46,0.44,0.70}
\definecolor{c4}{rgb}{0.91,0.16,0.54}
\definecolor{c5}{rgb}{0.40,0.65,0.12}
\definecolor{c6}{rgb}{0.90,0.67,0.01}

\newcommand{\solidline}[1][black]{\raisebox{2pt}{\tikz{\draw[-,color=#1,solid,line width = 0.5pt](0,0) -- (6mm,0);}}} 

\newcommand{\dashedline}[1][black]{\raisebox{2pt}{\tikz{\draw[-,color=#1,dashed,line width = 0.5pt](0,0) -- (6mm,0);}}}

\newcommand{\dashdotline}[1][black]{\raisebox{2pt}{\tikz{\draw[-,color=#1,dash dot,line width = 0.5pt](0,0) -- (6mm,0);}}}

\newcommand{\dottedline}[1][black]{\raisebox{2pt}{\tikz{\draw[-,color=#1,dotted,line width = 0.5pt](0,0) -- (6mm,0);}}}

\newcommand{\squareline}[1][black]{\raisebox{0pt}{\tikz{\draw[-,color=#1,solid,line width = 0.5pt](0.,0.8mm) -- (6mm,0.8mm); \node[rotate=0,scale=1,text=#1] at (3mm,0.8mm){\pgfuseplotmark{square*}}}}}

\makeatletter
\usetikzlibrary{calc}
\newcommand*\circled[1][]{\tikz[baseline=(char.base)]{
		\node[shape=circle,draw=black,fill=white,inner sep=0pt,line width=1pt,minimum height={\f@size*1.1},] (char) {\vphantom{WAH1g}{\textcolor{black}{\footnotesize{\textbf{#1}}}}};}}
\makeatother

\newcommand{\RomanNumeralCaps}[1]
\nolinenumbers

\title{Investigation of unsteady secondary flows and large-scale turbulence in heterogeneous turbulent boundary layers}

\author{D. D. Wangsawijaya\aff{1,2}
  \corresp{\email{D.D.Wangsawijaya@soton.ac.uk}}
 \and N. Hutchins\aff{1}}

\affiliation{\aff{1}Department of Mechanical Engineering, University of Melbourne, Victoria 3010, Australia
\aff{2}Aerodynamics and Flight Mechanics Research Group, Faculty of Engineering and Physical Sciences, University of Southampton SO17 1BJ, United Kingdom}

\begin{document}
\maketitle

\begin{abstract}
    Following the findings in \cite{wangsawijaya2020}, we re-examine the turbulent boundary layers developing over surfaces with spanwise heterogeneous roughness of various roughness wavelengths $0.32 \leq S/\overline{\delta} \leq 3.63$, where $S$ is the width of the roughness strips and $\overline{\delta}$ is the spanwise-averaged boundary-layer thickness. The heterogeneous cases induce counter-rotating secondary flows, and these are compared to the large-scale turbulent structures that occur naturally over the smooth wall. Both appear as meandering elongated high- and low-momentum streaks in the instantaneous flow field. Results here based on the triple decomposed velocity fluctuations suggest that the secondary flows might be spanwise-locked turbulent structures, with $S/\overline{\delta}$ governing the strength of the turbulent structures and possibly the efficacy of the surface in locking the structures in place (most effective when $S/\overline{\delta} \approx 1$). In terms of unsteadiness, we find an additional evidence from these conditional averages showing that the secondary flows exhibit maximum unsteadiness (or meandering) when $S/\overline{\delta} \approx 1$. Conditional averages of the fluctuating velocity fields of both spanwise heterogeneous and smooth wall cases result in structures that are strongly reminiscent of the streak-vortex instability model. Secondary flows and large-scale structures coexist in the limits where either $S/\overline{\delta} \gg 1$ or $S/\overline{\delta} \ll 1$, where the secondary flows scale on $\delta$ or $S$, respectively. When $S/\overline{\delta} \gg 1$, the secondary flows are locked about the roughness transition, while relatively unaltered large-scale structures occur further from the transition. In the case where $S/\overline{\delta} \ll 1$, $S$-scaled secondary flows are confined close to the surface, coexisting with unaltered larger scale turbulent structures that penetrate much deeper into the layer.
\end{abstract}


\section{Introduction}


We consider a specific case of heterogeneous roughness where the roughness varies in the spanwise direction. Spanwise heterogeneity is imposed by alternating rough and smooth strips to form a test surface over which a turbulent boundary layer is developed (figure~\ref{fig:setup_piv}). This type of roughness heterogeneity induces secondary flows in the form of counter-rotating streamwise rollers, which are apparent in the time-averaged velocity field. An examination of the instantaneous velocity field, however, reveals the secondary flows as elongated meandering high- and low-speed streaks, not unlike the large-scale/very-large-scale motions (LSM/VLSM) that occur naturally in wall-bounded turbulence \citep{wangsawijaya2020}.  

LSMs have commonly been associated with hairpin packets \citep{adrian2000,tomkins2003} and $\delta$-scaled meandering high- and low-momentum streaks \citep{hutchins2007a,hutchins2007b} in the outer layer of wall-bounded turbulent flows, where $\delta$ is either the boundary layer thickness, channel half-height, or pipe radius. In the logarithmic region, clusters of hairpin packets agglomerate to form VLSM or `superstructures' \citep{kim1999,guala2006,balakumar2007,hutchins2007a,dennis2011b,lee2011,hutchins2012,wu2012}. The imprint of symmetrical hairpin vortex packets is evident in certain statistical analyses of these structures. For example, two-point correlations of the fluctuating flow field exhibit an elongated low-momentum streak flanked by high-momentum streaks in the spanwise direction \citep{tomkins2003,ganapathisubramani2003,ganapathisubramani2005,hutchins2005,hutchins2007a,marusic2010,dennis2011a}. However, \cite{johansson1991} cautioned against this interpretation of the two-point correlation contours. Ensemble averaging and assumptions of spanwise homogeneity, as typically applied in the computation for smooth wall-bounded turbulence, enforce a plane of symmetry in the resulting conditional average. Asymmetrical large-scale structures (one-sided roll modes) have, in fact, been observed instantaneously in turbulent boundary layers formed over smooth walls \citep{lozano-duran2012,kevin2019b}, heterogeneous roughness \citep{vanderwel2019,wangsawijaya2020}, and converging-diverging (C-D) riblets \citep{kevin2017}. \cite{elsinga2010} observed that hairpin packets are most likely comprised of arch-like or cane-shaped structures. Similar to the near-wall cycle, an alternative model for the formation mechanism has been suggested for coherent structures in the outer layer. Large-scale elongated streaks with a sinuous instability and associated asymmetric staggered quasi-streamwise vortices have been observed in the logarithmic region and beyond \citep{flores2010,cossu2017,degiovanetti2017}, which are similar to, but at a much larger scale than, structures associated with the streak-vortex instability, which was initially developed as a model for near-wall streak formation \citep{jeong1997,waleffe2001,schoppa2002}. 

The naturally occurring large-scale structures and the secondary flows induced by spanwise heterogeneity share some similarities: both are characterised by roll modes and elongated high- and low-momentum streaks. An unsteadiness or streamwise periodic behaviour of the secondary flows have also been inferred from previous studies of spanwise heterogeneous roughness through the 1-D energy spectrograms \citep{nugroho2013,awasthi2018,medjnoun2018,zampiron2020,wangsawijaya2020}, two-point correlation maps \citep{kevin2017,kevin2019a,wangsawijaya2020}, and POD (proper orthogonal decomposition) of the turbulent fluctuation fields \citep{vanderwel2019}. It was suggested in \cite{wangsawijaya2020} that this unsteadiness is a function of the spanwise roughness wavelength $\Lambda = 2S$, where $S$ is the width of the roughness strip. The secondary flows strongly meander when $S/\delta \approx 1$, with a greater meandering amplitude as compared to the LSM/VLSM of smooth-wall turbulent flows and the secondary flows in the limiting cases where $S/\delta \gg 1$ or $S/\overline{\delta} \ll 1$. These observed similarities and differences pose questions about the nature of the secondary flows imposed by surface roughness in comparison to the large-scale structures in wall-bounded turbulence. The similarities might suggest that secondary flows and  large-scale structures share the same formation mechanism (a notion alluded to by \citealp{townsend1976}, p. 328--331). If that is the case, (i) is it possible that secondary flows are just phase-locked large-scale structures (and, large-scale structures are just non-phase-locked secondary flows)? Given the unsteadiness of secondary flows noted above, (ii) can this be explained as the natural meandering process of the large-scale structures? In \cite{chung2018}, the isovels of the mean streamwise velocity showed that in the limiting case scenarios where $S/\delta \gg 1$ or $S/\overline{\delta} \ll 1$ the secondary flows are confined within certain parts of turbulent boundary layer while other regions approach locally homogeneous conditions. This leads to the final question about (iii) the extent to which the secondary flows and the naturally present turbulent structures coexist in boundary layers formed over spanwise heterogeneous roughness. To answer these questions, we conduct an analysis of the fluctuating velocity components obtained from particle image velocimetry (PIV) measurements on turbulent boundary layers developing over surfaces composed of streamwise-aligned, spanwise-alternating sandpaper and cardboard strips (figure~\ref{fig:setup_piv}). The test surfaces cover a range of spanwise wavelength $0.32 \leq S/\overline{\delta} \leq 3.63$, where $\overline{\delta}$ is the spanwise-averaged boundary layer thickness at the measurement station.

\begin{figure}
	\centering
	\includegraphics[width=13cm, height=9.2cm, keepaspectratio]{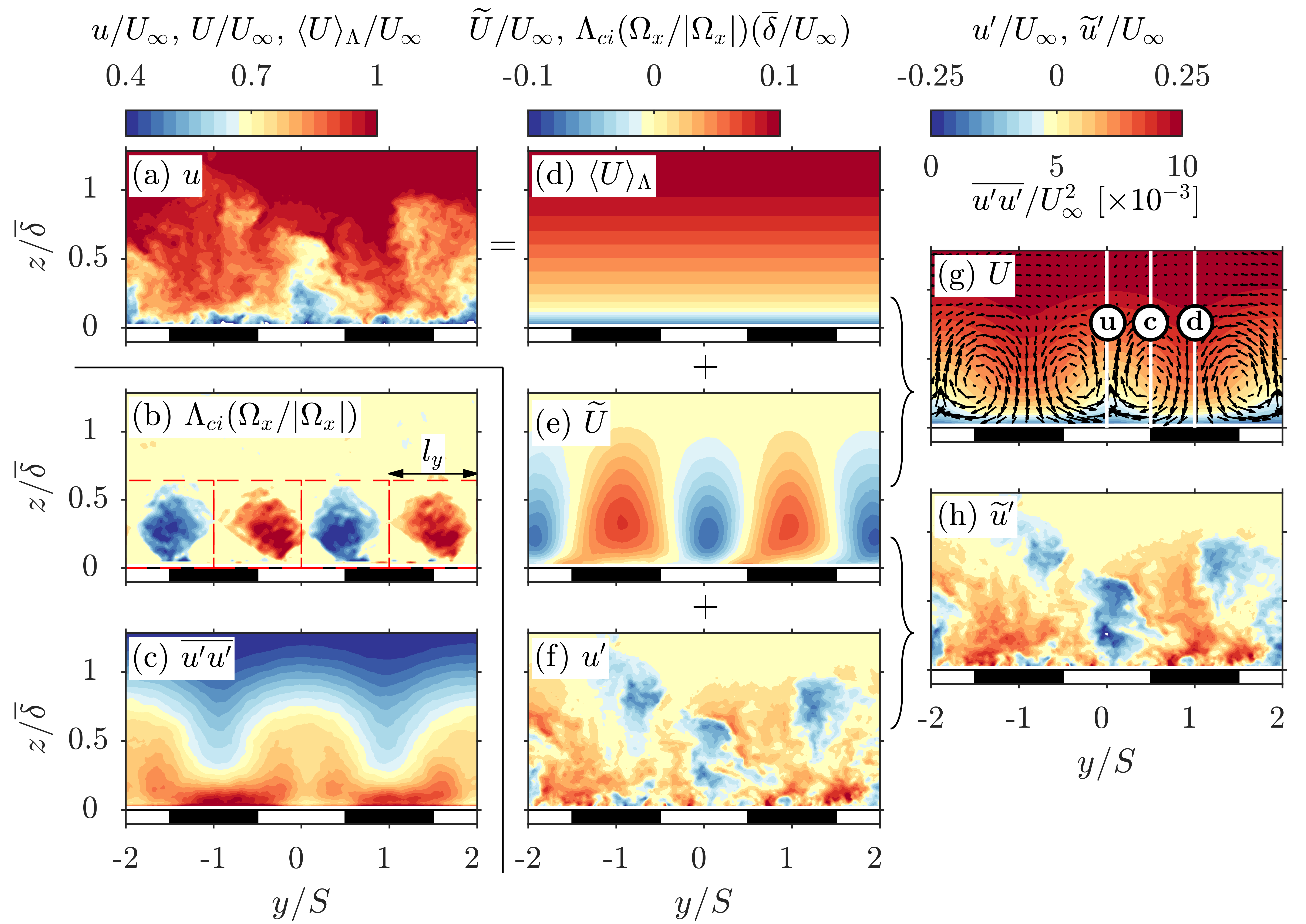}
	
	\caption{Triple-decomposition of (a) instantaneous snapshot streamwise velocity $u$ into (d) $yt$-average $\langle U \rangle_{\Lambda}$, (e) time-average, spatial fluctuation $\widetilde{U}$, and (f) instantaneous snapshot of turbulent fluctuation $u'$ for case SR50 ($S/\overline{\delta} = 0.62$). In (b), $\Lambda_{ci}$ is the mean swirl strength multiplied by the sign of streamwise vorticity $\Omega_x/|\Omega_x|$ and normalised by $\overline{\delta}/U_{\infty}$. Red dashed lines illustrate the extent of non-zero $\Lambda_{ci}$, whose width is $l_y$. (c) shows the variance of streamwise turbulent fluctuation $\overline{u'u'}$. (g) contours of time-averaged streamwise velocity $U$. Vectors indicate $V$ and $W$, downsampled for clarity. Solid white lines mark the three spanwise locations related to the mean secondary flows: common flow up (\circled[u]), centre of a mean secondary flow (\circled[c]), and common flow down (\circled[d]). Black and white patches indicate `rough' and `smooth' strips, respectively. (h) Contours of $\widetilde{u}'$, which is defined as $\widetilde{u}' \equiv \widetilde{U} + u'$.}
	\label{fig:U_decomposed}
\end{figure}

The axis system in this study $\boldsymbol{x} = (x, y, z)$ is defined as the streamwise, spanwise, and wall-normal directions, respectively, which corresponds to the instantaneous velocity components $\boldsymbol{u} = (u, v, w)$. As illustrated in figure~\ref{fig:U_decomposed}, $\boldsymbol{u}$ is triple decomposed into its temporal-, spatial-average, and the fluctuations \citep{raupach1982,finnigan2000,coceal2006},
\begin{eqnarray}
\boldsymbol{u}(y,z,t) & = & \boldsymbol{U}(y,z) + \boldsymbol{u'}(y,z,t) \label{eq:veldef1} \\	
& = & \boldsymbol{\langle U \rangle}_{\Lambda}(z) + \boldsymbol{\widetilde{U}}(y,z) + \boldsymbol{u'}(y,z,t) \label{eq:veldef2} \\  
& = & \boldsymbol{\langle U \rangle}_{\Lambda}(z) + \boldsymbol{\widetilde{u}'}(y,z,t) 
\label{eq:veldef3}
\end{eqnarray}
where $\boldsymbol{U}$ (figure~\ref{fig:U_decomposed}a) is the Reynolds (temporal) average, and $\boldsymbol{u'}$ (figure~\ref{fig:U_decomposed}f) is the turbulent fluctuations about this Reynolds average. $\boldsymbol{U}$ is further decomposed into its $yt$-average $\boldsymbol{\langle U \rangle}_{\Lambda}$ (figure~\ref{fig:U_decomposed}d) and the spatial fluctuations about this mean $\boldsymbol{\widetilde{U}}$ (figure~\ref{fig:U_decomposed}e). Since $\langle V \rangle_{\Lambda} = \langle W \rangle_{\Lambda} = 0$, $\boldsymbol{\widetilde{U}} = (\widetilde{U}, V, W)$. With this chosen decomposition method for spanwise heterogeneous roughness, $\boldsymbol{\widetilde{U}}$ can be considered as the stationary components of the secondary flows (or the dispersive components) and $\boldsymbol{u'}$ are both the advecting turbulence \emph{and} the unsteadiness of the secondary flows. We also introduce the quantity $\boldsymbol{\widetilde{u}'}$ (see \ref{eq:veldef3}), which is defined as $\boldsymbol{\widetilde{u}'} \equiv \boldsymbol{\widetilde{U}} + \boldsymbol{u'}$ (figure~\ref{fig:U_decomposed}h). It should be noted that for the reference smooth-wall case SW-2, as a result of spanwise homogeneity, $\boldsymbol{\widetilde{U}} = 0$ and $\boldsymbol{\widetilde{u}'} = \boldsymbol{u'}$. 

\section{Experimental set-up}

\subsection{Test surfaces}

\begin{table}
	\begin{center}
		\def~{\hphantom{0}}	
		\begin{tabular}{c c c c c c c c c c c c}
			\multicolumn{12}{c}{\textbf{Spanwise heterogeneous roughness}}\\[3pt]
			
			Case~~~ & $S$ & $\overline{\delta}$ & $S/\overline{\delta}$ & $x$ & $U_{\infty}$ & $Re_{x}$ & $Re_{\overline{\delta}}$ & $Re_{\overline{\theta}} $ & $\overline{\delta}^+$ & $z_{sheet}/\overline{\delta}$ & $z_{sheet}/S$ \\
			
			& [mm] & [mm] & & [m] & [ms$^{-1}$] & $[\times 10^{6}]$ & $[\times 10^{4}]$ & & & & \\[3pt]
			
			SR250-2 & 250 & 68.8 & 3.63 & 4.00 & 15.4 & 3.91 & 6.72 & ~9090 & -- & 0.49 & 0.13 \\
			SR160~~ & 160 & 70.0 & 2.28 & 4.00 & 15.3 & 3.91 & 6.84 & ~9200 & -- & 0.48 & 0.21 \\
			SR100~~ & 100 & 73.9 & 1.35 & 4.00 & 15.2 & 3.94 & 7.27 & 10150 & -- & 0.46 & 0.34 \\
			SR50~~~ & ~50 & 80.7 & 0.62 & 4.00 & 15.3 & 3.92 & 7.92 & 10220 & -- & 0.32 & 0.51 \\
			SR25~~~ & ~25 & 77.6 & 0.32 & 4.00 & 15.5 & 3.86 & 7.49 & ~9990 & -- & 0.18 & 0.55 \\
			
			& & & & & & & & & & &\\
			\multicolumn{12}{c}{\textbf{Reference smooth wall}}\\[3pt]
			
			Case~~~ & $S$ & $\delta_s$ & $S/\delta_s$ & $x$ & $U_{\infty}$ & $Re_{x}$ & $Re_{\delta_s}$ & $Re_{\theta_s} $ & $\delta_s^+$ & $z_{sheet}/\delta_s$ & $z_{sheet}/S$\\
			
			& [mm] & [mm] & & [m] & [ms$^{-1}$] & $[\times 10^{6}]$ & $[\times 10^{4}]$ & & & & \\[3pt]
			
			SW-2~~~ & -- & 56.4 & -- & 4.00 & 15.2 & 3.99 & 5.62 & 7320 & 2030 & 0.46 & -- \\
			
			SW-2~~~ & -- & 56.4 & -- & 4.00 & 15.2 & 3.99 & 5.62 & 7320 & 2030 & 0.24 & -- \\
			
		\end{tabular}
		\caption{Summary of spanwise heterogeneous roughness cases and the reference smooth-wall cases. $\overline{\delta}$ is the spanwise-averaged 98\% boundary-layer thickness of the surface with spanwise heterogeneity, while $\delta_s$ is the 98\% boundary-layer thickness of the reference smooth-wall case. $\overline{\theta}$ is the spanwise-averaged momentum thickness of the spanwise heterogeneous surfaces, while $\theta_s$ is the momentum thickness of the reference smooth-wall case. Reynolds number definitions are: $Re_x \equiv xU_{\infty}/\nu$, $Re_{\delta} \equiv \delta U_{\infty}/\nu$, $Re_{\theta} \equiv \theta U_{\infty}/\nu$, and $\delta^+ \equiv \delta U_{\tau}/\nu$, where $U_{\tau}$ is the friction velocity. $z_{sheet}$ is the wall-normal location of the wall-parallel PIV laser sheet, measured from the wall.}
		\label{tab:sr} 
	\end{center}
\end{table} 

The measurements are performed in an open return boundary layer wind tunnel in the Walter Basset Aerodynamics Laboratory at the University of Melbourne using the same experimental set-up and test surfaces as \cite{wangsawijaya2020}. The spanwise heterogeneous roughness (`SR') surfaces are constructed from spanwise-alternating strips of P-36 grit sandpaper (`rough' patch) and cardboard (`smooth' patch) of equal width $S$ (figure~\ref{fig:setup_piv}), with minimal variations in surface elevation between the two. Throughout this report, the rough patches are shaded black and the smooth strips are shaded white. This study considers measurements over heterogeneous surfaces of various $S$ at $x = 4$ m downstream of the sandpaper trip located at the inlet of the wind tunnel test section, covering a range of $0.32 \leq S/\overline{\delta} \leq 3.63$. For comparison to SR cases, measurements are also conducted over a reference smooth-wall (`SW') case at the same $Re_x \equiv xU_{\infty}/\nu$ as the SR cases, where $U_{\infty}$ is the freestream velocity. The details of spanwise heterogeneous and smooth wall reference cases are summarised in table~\ref{tab:sr}. Note that the `-2' suffix in SR250 and SW cases refer to measurements at $x = 4$ m, such that it is consistent with the nomenclature in \cite{wangsawijaya2020}. 

\subsection{Particle image velocimetry (PIV)}

\begin{figure}
	\centering
	\includegraphics[width=14.5cm, height=5cm, keepaspectratio]{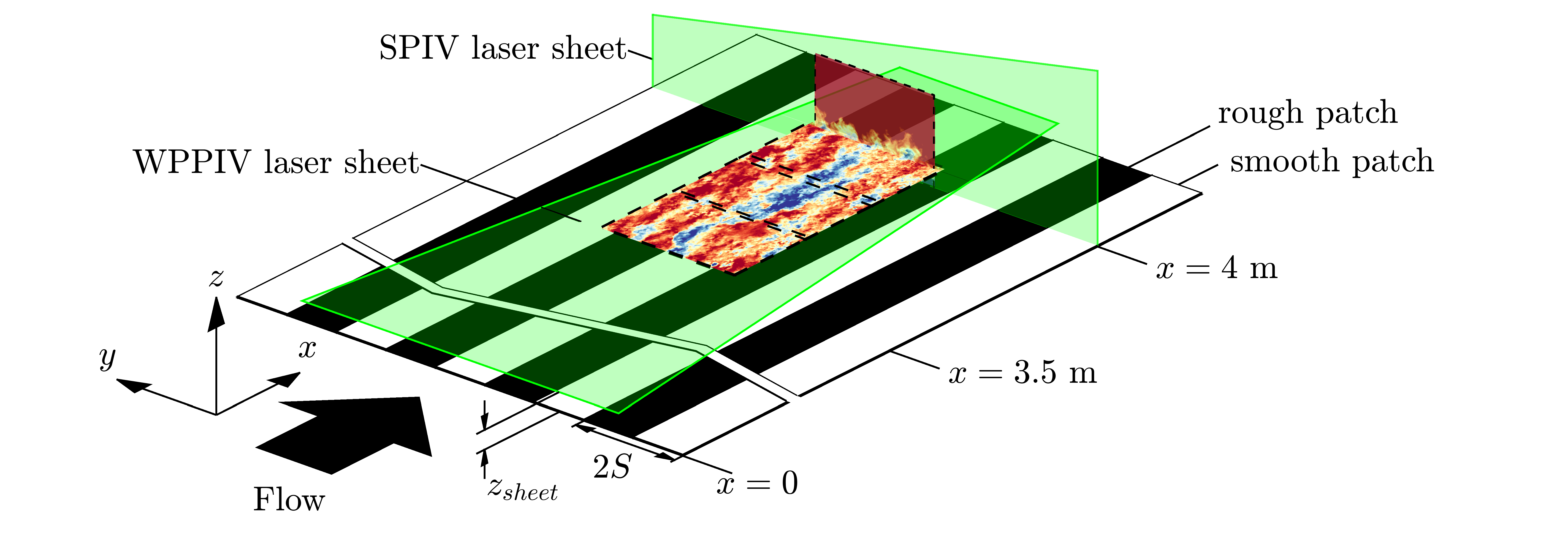}	
	\caption{Schematic of PIV experimental set-up: stereoscopic PIV (SPIV) in $yz$-plane and wall-parallel PIV (WPPIV) in plane $xy$-plane. Measurements in these planes are not conducted simultaneously.}
	\label{fig:setup_piv}
\end{figure} 

PIV measurements are conducted over the test surfaces in two planes: stereoscopic PIV (SPIV) in the cross-stream ($y$--$z$) plane and wall-parallel PIV (WPPIV) in the streamwise-spanwise ($x$--$y$) plane (figure~\ref{fig:setup_piv}). Measurements in both planes are non-time-resolved and non-simultaneous. The SPIV plane is located at $x = 4$ m downstream of the trip, coinciding with the WPPIV plane which spans $3.5 \leq x \leq 4.01$ m. SPIV images are captured by two pco.4000 cameras in forward scatter arrangement, which gives the field of view (FOV) of $4\delta_s \times 3\delta_s$ ($y \times z$). For WPPIV, the FOV is obtained by stitching images from three pco.4000 cameras, resulting in a FOV of total size $9\delta_s \times 5\delta_s$ ($x \times y$). The wall-normal location of the WPPIV laser sheet relative to the wall $z_{sheet}$ varies between SR cases (see table~\ref{tab:sr}), depending on whether the secondary flow sizes are governed by $\overline{\delta}$ or $S$ (see equation 3.1 in \citealp{wangsawijaya2020}). For the cases where $S/\overline{\delta} \geq 1$ (SR250-2, SR160, and SR100), the sheet is located at $z_{sheet}/\overline{\delta} \approx 0.5$, while for $S/\overline{\delta} < 1$ (SR50 and SR25), the sheet is located at $z_{sheet}/S \approx 0.5$. To accommodate the variation of $z_{sheet}$, WPPIV measurements for the reference smooth-wall case SW-2 are conducted in two $xy$-planes, at $z_{sheet}/\delta_s = 0.46$ and 0.24. It should be noted that for SR250-2 and SR160 cases, only half of the spanwise roughness wavelength is captured in the SPIV and WPPIV images due to the limited FOV width. The complete description of the SPIV and WPPIV experimental set-ups are given in the appendix A and B of \cite{wangsawijaya2020}.     

\section{Meandering of secondary flows}
\label{sub:sond_eq1}

In \cite{wangsawijaya2020}, meandering of the low-speed streaks, presumably the manifestation of the secondary flows in the instantaneous flow field, is inferred from the $y$--$z$ plane by a spanwise leaning behaviour and asymmetry of the low-speed features. That is, the structures lean sideways (in $y$) depending on the sign of $v'$, with the intermediate cases ($S/\overline{\delta} \approx 1$) showing the strongest leaning amplitude (figure 18 in \citealp{wangsawijaya2020}). In the wall-parallel ($x$--$y$) plane, meandering was implied by the two-point correlations of $u'$, showing a strong periodicity in the cases where $S/\overline{\delta} \approx 1$ when computed at certain $y$ locations (figure 19 in \citealp{wangsawijaya2020}). In the current study, analysis of the WPPIV data (for the reference smooth wall case and spanwise heterogeneous roughness) is extended to permit further examination of this meandering behaviour. Further analysis of the similarities (and differences) between the secondary flows and naturally occurring LSMs/VLSMs in homogeneous canonical flows is also possible.

\begin{figure}
	\centering
	\includegraphics[height=11cm, width=13cm, keepaspectratio]{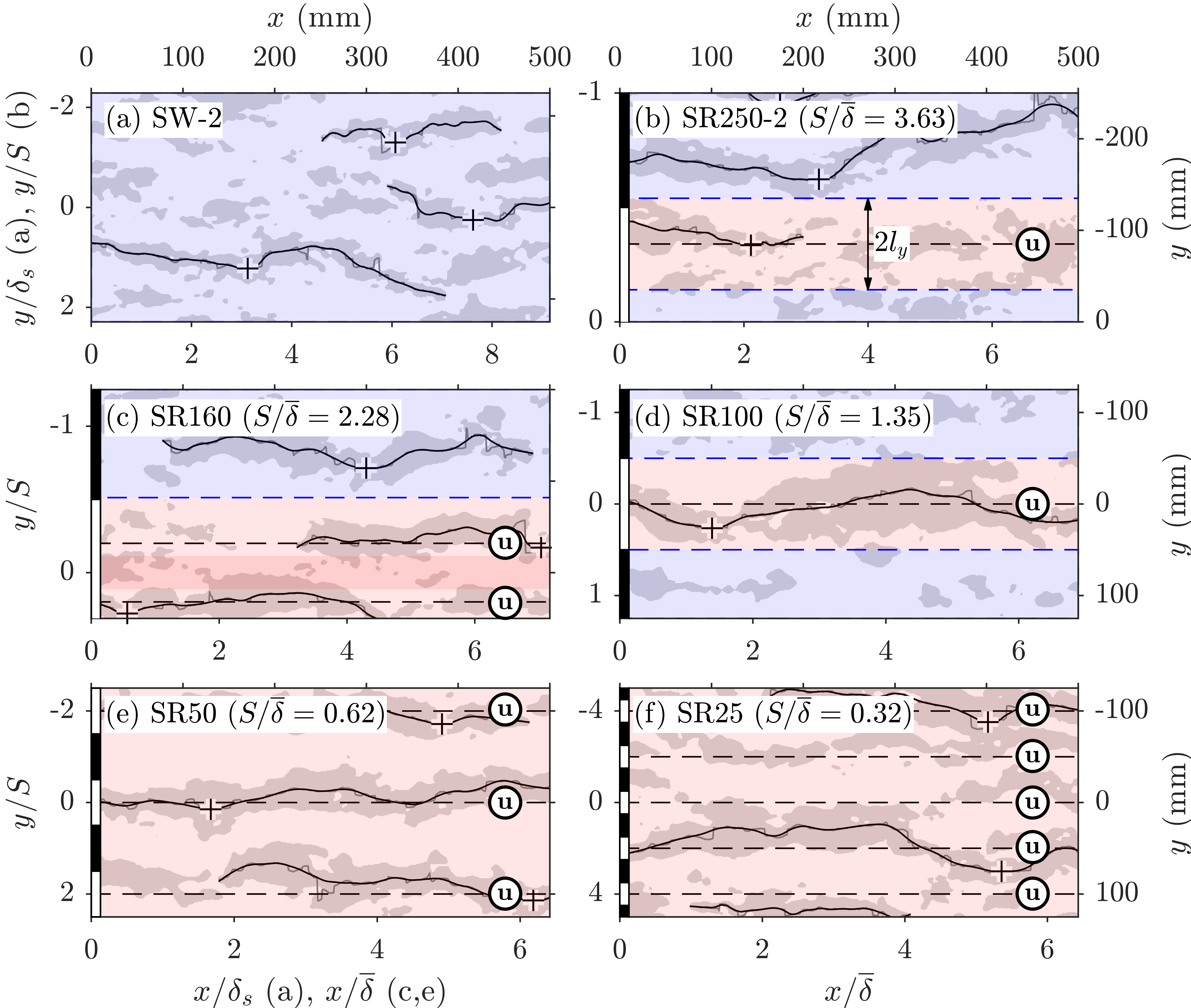}
	\caption{Detected low-speed structures for (a) reference case SW-2 at $z_{sheet}/\delta_s = 0.46$ and SR cases: (b) SR250-2, (c) SR160, (d) SR100 ($z/\overline{\delta} = 0.46$), (e) SR50, (f) SR25, about the center of the secondary flows. Grey coloured contours are the low-speed structures, $\widetilde{u}'/U_{\infty} < -0.03$. `+' marks the minima (in terms of y location) of a low-speed structure. The spines of the detected low-speed structures are shown in solid lines (from PIV data: \solidline[LightGray], smoothed: \solidline[black]). Dashed lines (\dashedline[black]) are the spanwise locations of the common flow up of the secondary flows (marked by \circled[u], see figure~\ref{fig:U_decomposed}g for these locations in the $yz$-plane). The low-speed structures related to the secondary flows due to spanwise heterogeneity are assumed to occur inside the red-shaded area, spanning $2l_y$ about the location of common flow up, as shown in (b) and figure~\ref{fig:swirl}. In (b--f), white and black patches illustrate the arrangement of smooth and rough strips, respectively, underneath the WPPIV planes.}
	\label{fig:spines}
\end{figure}

\begin{figure}
	\centering
	\includegraphics[height=4.2cm, width=13cm, keepaspectratio]{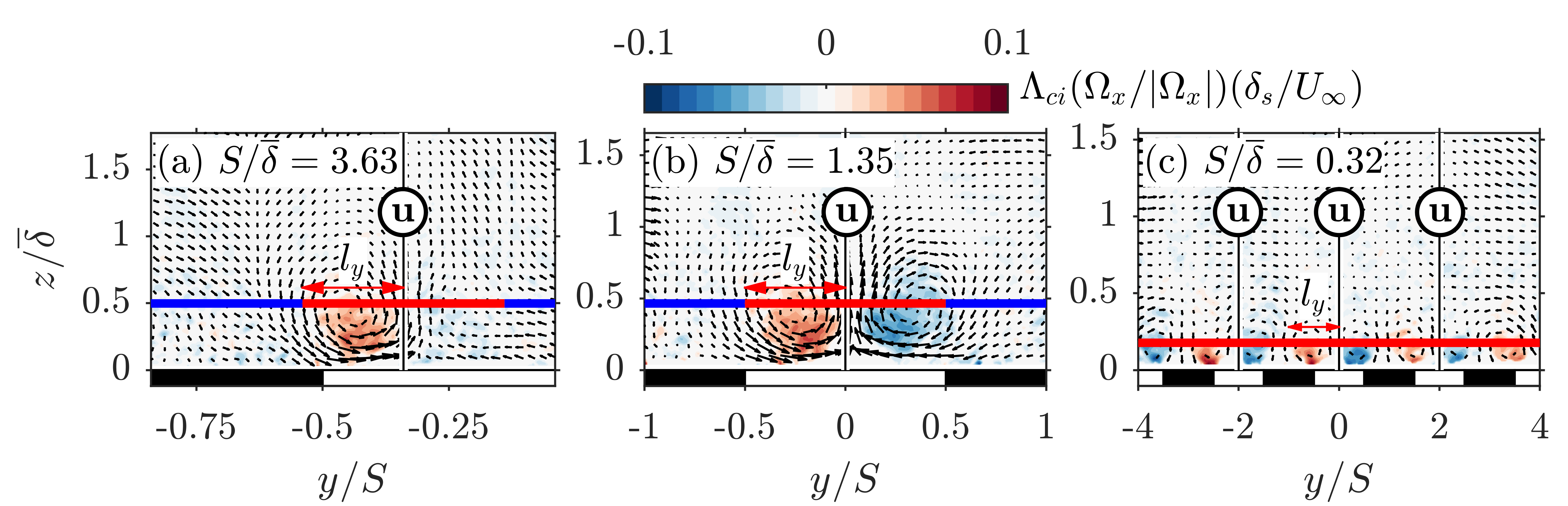}
	\caption{Contours of mean swirl strength $\Lambda_{ci}$ in $yz$-plane multiplied by the sign of vorticity $\Omega_x/|\Omega_x|$ for case: (a) SR250-2 ($S/\overline{\delta} = 3.63$), (b) SR100 ($S/\overline{\delta} = 1.35$), (a) SR25 ($S/\overline{\delta} = 0.32$). Contours are normalised by $\delta_s/U_{\infty}$. Vectors indicate $V$ and $W$, downsampled for clarity. \circled[u] is the location of common flow up. Red (\solidline[red]) and blue (\solidline[blue]) solid lines illustrate the WPPIV laser sheets ($xy$-plane in figure~\ref{fig:spines}) of each case. Secondary flows are assumed to occur along the red lines, LSM/VLSM along the blue lines. Red arrows indicate $l_y$, the spanwise extent of the mean secondary flows.}
	\label{fig:swirl}
\end{figure}

We attempt to reconstruct the meandering of large-scale structures and secondary flows through conditional averaging of the fluctuating velocity field. Figure~\ref{fig:spines} shows the instantaneous low-speed structures from a single representative snapshot taken from the WPPIV measurements for the reference smooth wall case SW-2 (figure~\ref{fig:spines}a) at $z/\delta_s = 0.46$ and all SR cases (figure~\ref{fig:spines}b--f) at a wall-normal location that corresponds to the approximate centre of the roll-modes associated with the mean secondary flows. Gray coloured contours show negative fluctuations of $\widetilde{u}' \equiv \widetilde{U} + u'$ (total velocity with the global $yt$-average subtracted), $\widetilde{u}'/U_{\infty} < -0.03$. To limit the analysis to long, large-scale structures, the velocity field is filtered with a box filter of size $0.1\delta_s \times 0.1\delta_s$ for SW-2 and $0.1\overline{\delta} \times 0.1\overline{\delta}$ for SR cases and only structures with length $\geq 3\overline{\delta}$ ($\geq 3\delta_s$ for SW-2) are retained for analysis. The `spine' of each detected low-speed region is constructed from the spanwise midpoint of the structure at all streamwise locations along the length of the detected feature (gray solid lines, \solidline[LightGray] in figure~\ref{fig:spines}). This `spine' is further smoothed with a 1-D low-pass filter whose length is $\overline{\delta}$ ($\delta_s$ for SW-2), shown by the black solid lines (\solidline[black]) in figure~\ref{fig:spines}. Conditional averaging of the turbulent fluctuation $u'$ (instead of $\widetilde{u}'$) is computed at the minima (in $y$) of the smooth spines fitted to the detected low-speed structures, as marked by the `+' symbols in figure~\ref{fig:spines}.

Figure~\ref{fig:spines}(d,e) also shows how the meandering of low-speed structures is clearly `phase-locked' about the spanwise location of the common flow up of the mean secondary flows (marked by dashed lines, \dashedline[black] and \circled[u]). This is expected since $\widetilde{u'}$ and $\widetilde{U}$ is phase-locked. Ensemble averaging of the total velocity field, as shown in the contours of swirl strength $\Lambda_{ci}$ (figure~\ref{fig:swirl}), reveal time-averaged large-scale secondary flows, even for cases where $S/\overline{\delta} \gg 1$ (figure~\ref{fig:swirl}a) and $S/\overline{\delta} \ll 1$ (figure~\ref{fig:swirl}c). Based on the time-averaged secondary flows depicted in figure~\ref{fig:swirl}, it can be assumed that the secondary flows in the instantaneous velocity fields meander about a certain spanwise location for all $S/\overline{\delta}$ cases. Here, it is assumed that the secondary flows due to spanwise heterogeneity cause low-speed streaks to meander about $y_u$, the spanwise location of common flow up (\circled[u] in figure~\ref{fig:spines} and~\ref{fig:swirl}), spanning the area shaded by red in figure~\ref{fig:spines}. This area spans $y_u \pm l_y$ (figure~\ref{fig:spines}b), where $l_y$ is the width of the mean secondary flow roll-mode, as illustrated in figure~\ref{fig:U_decomposed}(b) and~\ref{fig:swirl}. Under this assumption, all detected low-speed structures could be classified as either secondary flows due to spanwise heterogeneity (red shaded area) or LSM/VLSM over homogeneous roughness regions in the cases where $S/\overline{\delta} > 1$ (blue shaded area in figure~\ref{fig:spines}b--d). In the cases where $S/\overline{\delta} < 1$ (figure~\ref{fig:spines}e,f), $l_y = S$ and all detected low speed structures will be categorized as belonging to the secondary flows; hence this categorisation is somewhat flawed. It should be noted that although the secondary flows fill the entire spanwise extent of the turbulent boundary layers in these cases, larger structures whose scale is $\overline{\delta}$ also coexist with the secondary flows. However, these two cannot be distinguished in cases where $S/\overline{\delta} \leq 1$ because the currently available WPPIV snapshots in the $x$--$y$ plane are obtained at a $z_{sheet}$ height that is centred on the roll modes and also because the method used to separate secondary flows and LSM/VLSM is based only on the spanwise location of the secondary flows (and not the spanwise lengthscale or wall-normal extent of the detected structure). A method to discriminate secondary flows from large scale structures for the case where $S/\overline{\delta} \ll 1$ based on POD (proper orthogonal decomposition) filtering is proposed in \S\ref{sub:sond_ll1}. The locations of the wall-parallel slices relative to the secondary flows and the locations of red- and blue-shaded regions in $y$--$z$ plane are illustrated in figure~\ref{fig:swirl}. For all subsequent conditional averaging, features are assigned to the red or the blue regions based on which region the detected y-minima of the low-speed structures (which is the condition vector) resides.  


\begin{figure}
	\centering
	\includegraphics[height=10.25cm, width=13cm, keepaspectratio]{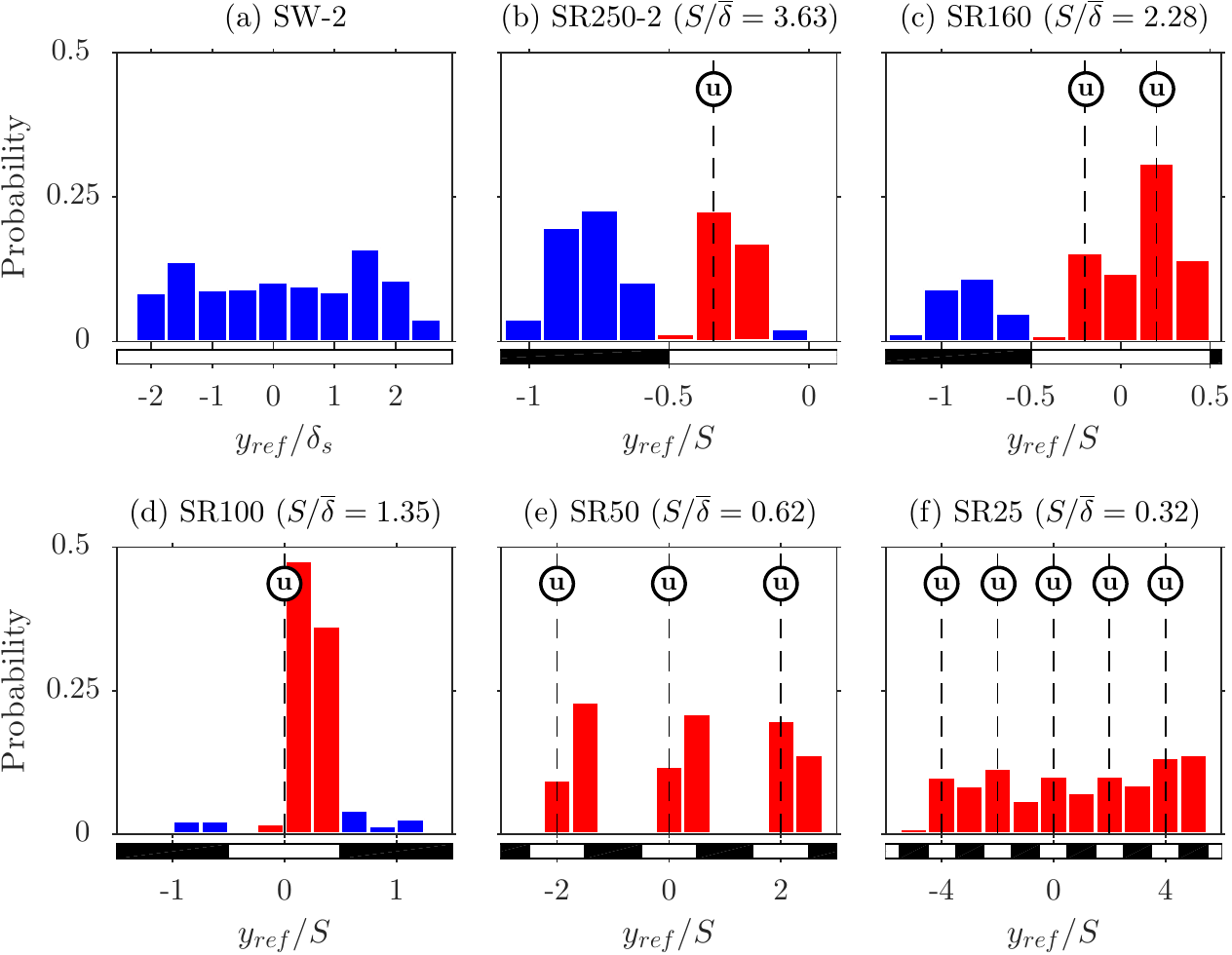}
	\caption{Histogram of the spanwise location of the detected minima of low-speed structures $y_{ref}$ for (a) reference case SW-2 at $z/\delta_s = 0.46$ and SR cases: (b) SR250-2, (c) SR160, (d) SR100, (e) SR50, (f) SR25 about the center of the secondary flows. Red bars (\textcolor{red}{$\blacksquare$}) indicate $y_{ref}$ detected inside the red-shaded area in figure~\ref{fig:spines}, while blue bars (\textcolor{blue}{$\blacksquare$}) are detected inside the blue-shaded area in the same figure. Dashed lines (\dashedline[black]) are the spanwise locations of the common flow up of the mean secondary flows (marked by \circled[u]).}
	\label{fig:hist_yref}
\end{figure} 

Figure~\ref{fig:hist_yref} shows the histogram of the detected low speed streaks (identified in the manner depicted in figure~\ref{fig:spines}) for all SR cases, split into those that we crudely classified as secondary flows (occuring within the red regions of figure~\ref{fig:spines}) and LSM/VLSM (occurring within the blue regions). The histogram counts the observed $y_{ref}$ minima (denoted by the black $+$ sign in figure~\ref{fig:spines}) in these two regions. The probability of the occurrence of long, low-speed structures are distributed equally across the span of the FOV in the reference smooth wall case SW-2 (figure~\ref{fig:hist_yref}a). For the case SR250-2 ($S/\overline{\delta} = 3.63$, red bars in figure~\ref{fig:hist_yref}b) and SR160 ($S/\overline{\delta} = 2.28$, figure~\ref{fig:hist_yref}c), the low-speed structures related to the secondary flows comprise 41\% and 74\% of all structures detected across the FOV, respectively (as a reference the red shaded regions associated with the secondary flows consist of 38\% and 53\% respectively of the available total area in these cases). The secondary flows dominate as $S/\overline{\delta}$ approaches 1 (86\% of detections occur within the red regions for case SR100, which occupy 40\% of the total area, see figure~\ref{fig:hist_yref}d), and they are also clearly `phase-locked' to the location of common flow up (case SR50, figure~\ref{fig:hist_yref}e). For the smallest $S/\overline{\delta}$ case (case SR25, figure~\ref{fig:hist_yref}f), the spines are more evenly distributed across the span compared to cases SR100 and SR50 (figure~\ref{fig:hist_yref}d,e), with a hint of some residual spanwise locking of the structures (higher probabilities over common flow up regions). 

\begin{figure}
	\centering
	\includegraphics[height=5.9cm, width=12.6cm, keepaspectratio]{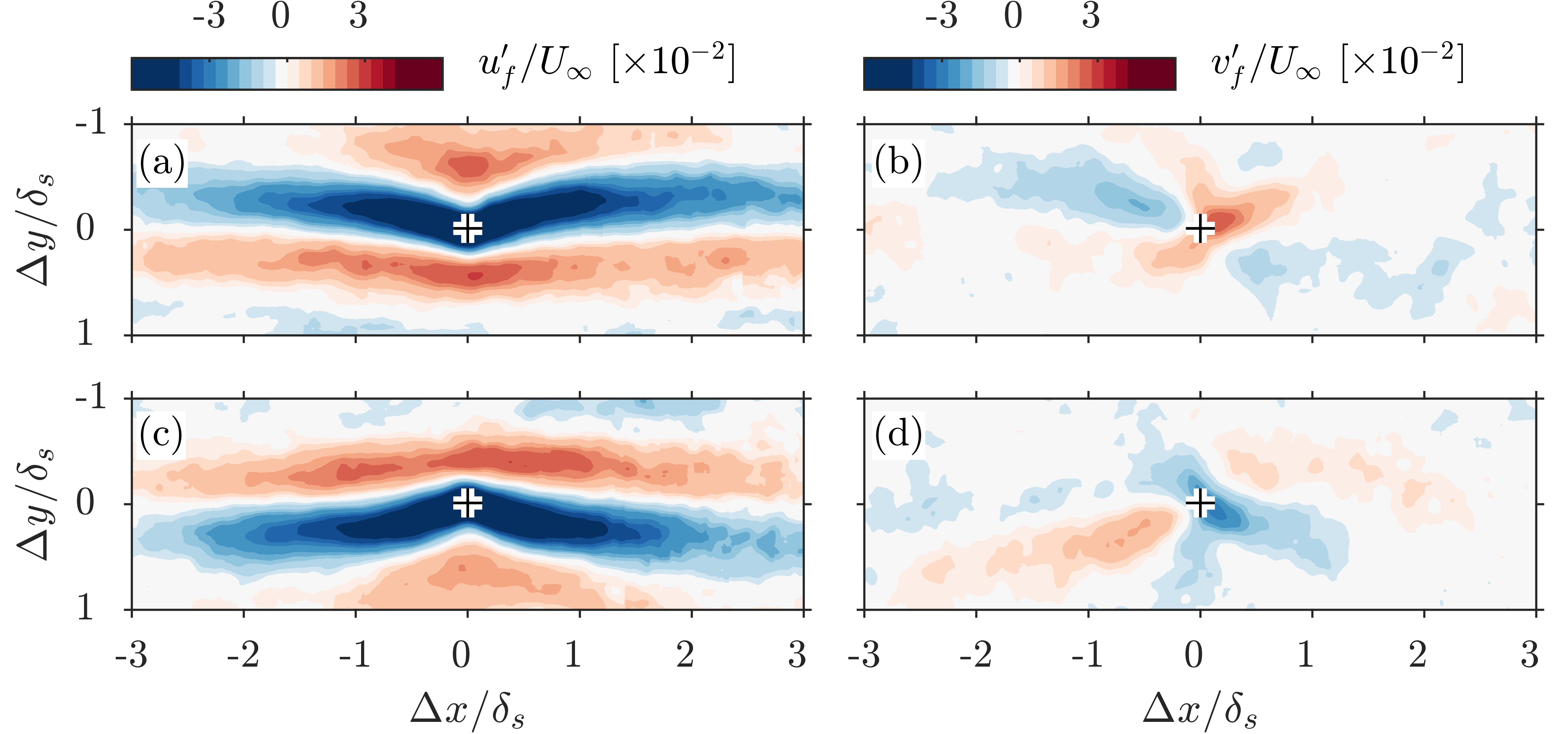}
	\caption{Contours of filtered turbulent fluctuation (a,c) $u_f'$ and (b,d) $v_f'$ conditionally averaged at (a,b) the minima and (c,d) maxima of the detected low-speed structures illustrated in figure~\ref{fig:spines} for the reference case SW-2 at $z/\delta_s = 0.46$.}
	\label{fig:condave_uv}
\end{figure}

Figure~\ref{fig:condave_uv} shows the conditional average of filtered streamwise $u_f'$ (figure~\ref{fig:condave_uv}a,c) and spanwise $v_f'$ (figure~\ref{fig:condave_uv}b,d) turbulent fluctuations for the reference smooth wall case SW-2 at $z/\overline{\delta_s} = 0.46$. The subscript `$f$' denotes the velocity fields filtered with a box filter of $0.1\delta_s \times 0.1\delta_s$ size ($0.1\overline{\delta} \times 0.1\overline{\delta}$ for SR cases). Figure~\ref{fig:condave_uv}(a,b) shows $u_f'$ conditionally averaged at the local $y$ minima of the spines of the detected low-speed structures as illustrated in figure~\ref{fig:spines}(a), while figure~\ref{fig:condave_uv}(c,d) is conditionally averaged at the maxima of the detected low-speed structures (not shown in figure~\ref{fig:spines}a). The contours of $u_f'$ show a low-speed structure flanked by two high-speed structures meandering to the left and right depending on the reference location (minima or maxima) where the conditional average is computed. The meandering tendency is also shown in $v_f'$, where the conditional average at the minima of low-speed structures largely correspond to $v_f' > 0$ (figure~\ref{fig:spines}b) and the maxima to $v_f' < 0$ (figure~\ref{fig:spines}d). Diagonal alignment of $v'$, similar to that observed in \cite{sillero2014} and \cite{desilva2018}, is also apparent in these contours. 

To obtain a complete picture of how the secondary flow meanders, similar conditional averaging is also computed in the $yz$-plane from SPIV measurements. Since the minima and maxima of the spine-fitted low-speed structure cannot be observed in the $yz$-plane, a different conditional averaging approach must be taken to detect these same events. Informed by the conditional averages shown in figure~\ref{fig:condave_uv}, we elect to use the a simultaneous detection criteria of negative $\widetilde{u}'$ and either $v' > 0$ or $< 0$ to approximately detect the minima and maxima of the fitted structure spines (cases shown in figure~\ref{fig:condave_uv}a,b and c,d respectively). The conditional average is computed at the common flow up (\circled[u] in figure~\ref{fig:U_decomposed}g) in $y$ and at $z/\overline{\delta} = 0.1$, as close to the wall as the FOV permits. For the reference case SW-2, the average is computed at any point in $y$ along $z/\delta_s = 0.1$ where the conditions are satisfied. The conditions $\widetilde{u}' < 0$, $v' > 0$ and $\widetilde{u}' < 0$, $v' < 0$ are each satisfied for 25\% of the smooth wall realisations. The velocity fields in the $yz$-plane are filtered with a box filter of size $0.1\delta_s \times 0.1\delta_s$ ($0.1\overline{\delta} \times 0.1\overline{\delta}$ for SR cases). It should be highlighted that these condition vectors for the ensemble averaging differ between the $x$--$y$ and $y$--$z$ plane and are only intended to show the representation of the high- and low-speed structures to complement the conditional averages in the $x$--$y$ plane. As a direct comparison, the conditional average at the exact same condition points in $x$--$y$ and $y$--$z$ plane has also been computed, showing a good agreement between both planes (not shown here for brevity, but available in \citealp{wangsawijaya2020-PHD}, p. 157--158).

\begin{figure}
	\centering
	\includegraphics[height=17.1cm, width=13cm, keepaspectratio]{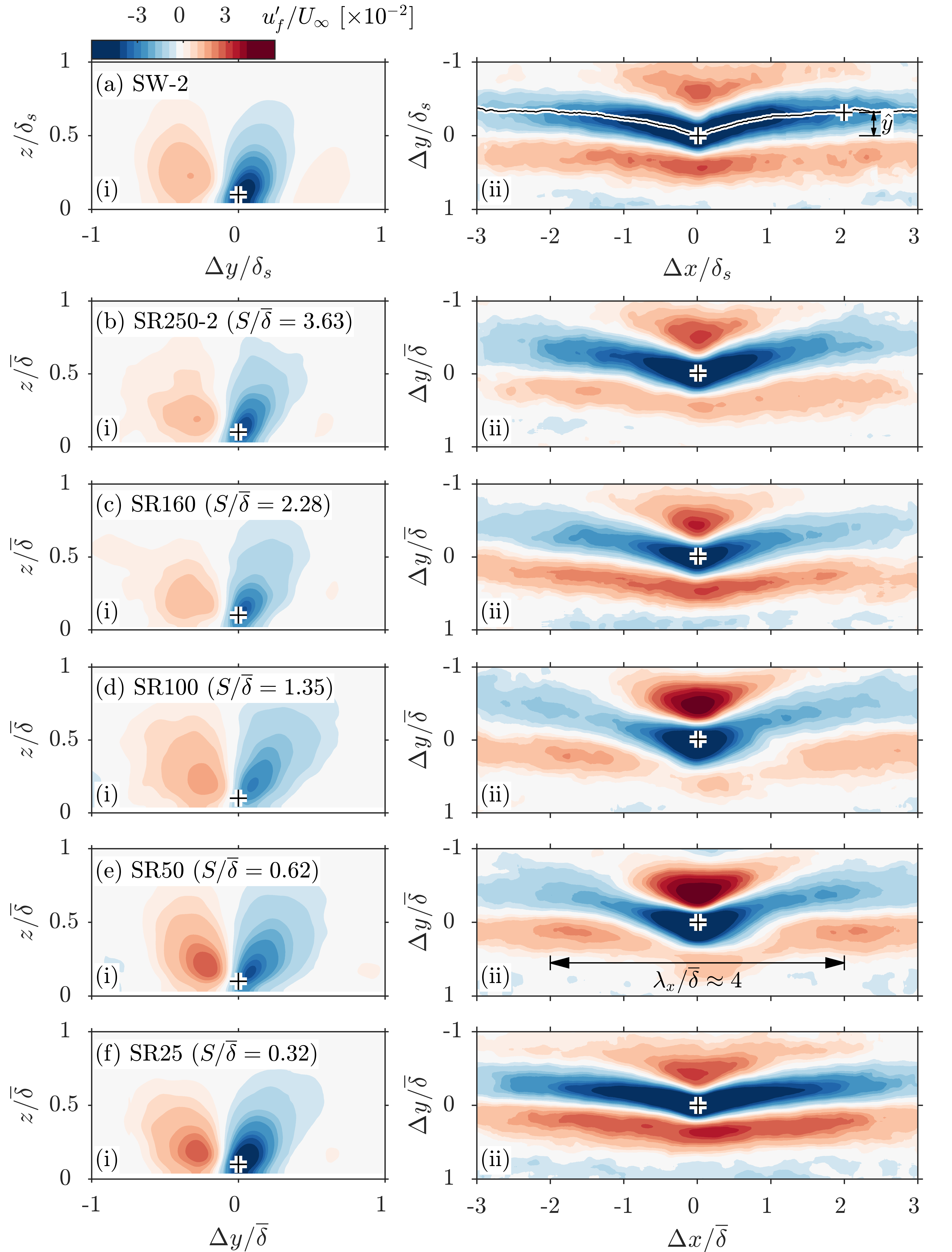}
	\caption{Contours of conditionally averaged turbulent fluctuation $u'$ for (a) reference case SW-2 at $z/\delta_s = 0.46$ and all SR cases: (b) SR250-2, (c) SR160, (d) SR100, (e) SR50 and (f) SR25. Conditions: (i) $\widetilde{u}' < 0$ and $v' > 0$ at the $y$ location of common flow up (\circled[u] in figure~\ref{fig:spines}), (ii) minima of detected low-speed structures ($\widetilde{u}'/U_{\infty} < -0.03$) assumed to be the secondary flows (red-shaded regions in figure~\ref{fig:spines}). In (a)(ii), solid black line (\solidline[black]) is the `spine' extracted from the conditionally averaged low-speed structure and $\hat{y}$ is the amplitude of meandering.}
	\label{fig:condave_eq1}
\end{figure} 

Conditionally averaged $u_f'$ for the reference smooth wall case SW-2 and all SR cases computed for features within the red-shaded region in figure~\ref{fig:spines} are shown in figure~\ref{fig:condave_eq1}. The plots show a low-speed structure flanked by two high-speed structures in both the $yz$- (i) and $xy$-planes (ii). Subplot (i) shows the tendency of the structures in the cross-plane to lean sideways to the right when $v_f' > 0$, corresponding to the minima of $y_{ref}$ in the wall-parallel plane (ii). The left-leaning tendency in the cross-plane for $v_f' < 0$, corresponding to the maxima of $y_{ref}$ in the wall-parallel plane, is not shown for brevity. The leaning is strongly one-sided for the cases where $S/\overline{\delta} \approx 1$ in figure~\ref{fig:condave_eq1}(d,e)(i), where the low-speed structures lean to one side and are flanked by asymmetric high-speed structures which are highly one-sided. For the reference smooth wall case SW-2 in figure~\ref{fig:condave_eq1}(a) and the heterogeneous cases where $S/\overline{\delta} \gg 1$ (figure~\ref{fig:condave_eq1}b,c), the sideways leaning of the low-speed event is less prominent with a more symmetric arrangement of flanking high-speed events. The conditionally averaged velocity fields in the $xy$-plane (ii) further confirm the differences between the smooth wall and the limiting cases compared to the $S/\overline{\delta} \approx 1$ cases. In the wall-parallel plane, the structures exhibit stronger meandering and asymmetry in figure~\ref{fig:condave_eq1}(d,e)(ii) when $S \rightarrow \overline{\delta}$ compared to the reference smooth wall and the limiting cases where $S \gg \overline{\delta}$ or $S \ll \overline{\delta}$ in figure~\ref{fig:condave_eq1}(a--c,f). The shape of these structures provide additional evidence for the meandering inferred from the contours of normalised two-point correlation of $u'$ in figure 19(e,f) in \cite{wangsawijaya2020}, where a clear streamwise anti-phase pattern was exhibited by the $S/\overline{\delta} \approx 1$ cases. The approximate streamwise wavelength observed in figure~\ref{fig:condave_eq1} is $\lambda_x/\overline{\delta} \approx 4$ (see figure~\ref{fig:condave_eq1}(e)ii), similar to that observed in the contours of two-point correlation coefficients and spectrograms from hot-wire anemometry measurements for the same surfaces in \cite{wangsawijaya2020}. Structures resulting from this conditional average for the limiting cases where $S/\overline{\delta} \gg 1$ and $S/\overline{\delta} \ll 1$ are further discussed in \S\ref{sub:sond_gg1} and \S\ref{sub:sond_ll1}, respectively.

\begin{figure}
	\centering
	\includegraphics[height=4.25cm, width=10.85cm, keepaspectratio]{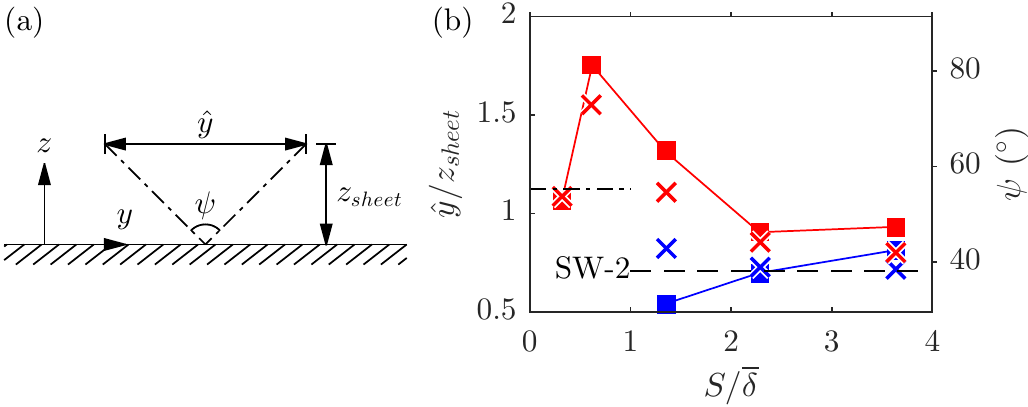}
	\caption{(a) Illustration of meandering (spanwise leaning) angle $\psi$ in $y$--$z$ plane, constructed from meandering amplitude $\hat{y}$ and the wall-normal location of the WPPIV laser sheet $z_{sheet}$. (b) Meandering amplitude $\hat{y}$ measured from the `spines' of the conditionally averaged structures (see figure~\ref{fig:condave_eq1}(a)ii) and normalised by $z_{sheet}$ as a function of $S/\overline{\delta}$. \squareline[blue]: $\hat{y}$ measured from the structures assumed to be LSM, when available (blue-shaded regions in figure~\ref{fig:spines}) and \squareline[red]: $\hat{y}$ measured from the structures assumed to be secondary flows (red-shaded regions in figure~\ref{fig:spines}). Measured $\hat{y}$ of the reference smooth-wall case SW-2, \dashdotline: $z_{sheet}/\delta_s = 0.24$ and \dashedline: $z_{sheet}/\delta_s = 0.46$. \textcolor{blue}{$\times$}: $\hat{y}$ of structures assumed to be LSM, \textcolor{red}{$\times$}: $\hat{y}$ of structures assumed to be secondary flows, both extracted from $u'$ instead of $\widetilde{u}'$ (see appendix~\ref{app} for details).}
	\label{fig:yhat}
\end{figure}

We next attempt to quantify the meandering of the conditionally averaged structures in the $x$--$y$ plane. This will supplement the quantification of the leaning angle in the $y$--$z$ plane conducted in \cite{wangsawijaya2020}. `Spines', similar to those fitted to the instantaneous low-speed structures in figure~\ref{fig:spines}, are now fitted to the conditionally averaged low-speed structures in figure~\ref{fig:condave_eq1}. The meandering amplitude $\hat{y}$ is defined as the average distance between the $y$ location of the spines at $\Updelta x/\overline{\delta} = 0$ and $\Updelta x/\overline{\delta} \pm 2$ (see figure~\ref{fig:condave_eq1}(a)ii). It should be noted that the wall-normal location of the $x$--$y$ plane (i.e. the WPPIV laser sheet location $z_{sheet}$) differs between cases, depending on the size of the secondary flows (table~\ref{tab:sr}) -- hence the measured $\hat{y}$ is not directly comparable between SR and SW cases. Alternatively, meandering can also be quantified with the spanwise leaning angle in the $y$--$z$ plane, $\psi \equiv 2 \tan^{-1}(\hat{y}/2 z_{sheet})$ (figure~\ref{fig:yhat}a). It should be noted that the definitions of $\hat{y}$ and $\psi$ are similar to those extracted from the conditionally averaged structures in the $y$--$z$ plane for \cite{wangsawijaya2020} (cf. figure 17e and 18). 

Figure~\ref{fig:yhat}(b) shows $\hat{y}$ normalised by $z_{sheet}$ (left hand side abscissa) and $\psi$ (right hand side abscissa) of the secondary flows (\squareline[red]) as a function of $S/\overline{\delta}$ ($z_{sheet}$ is roughly the radius of the secondary flows, see figure~\ref{fig:swirl}). The red symbols on this figure show the meandering amplitude of the secondary flows as a function of $S/\overline{\delta}$, and clearly this amplitude is maximum at $S/\overline{\delta} = 0.62$ (SR50, $S/\overline{\delta} \approx 1$), similar to the trend previously reported for the $y$--$z$ plane (figure 18 in \citealp{wangsawijaya2020}). The amplitude of the meandering normalised by the wall-normal sheet location $\hat{y}/z_{sheet}$ is approximately equal to or less than 1 for the smooth wall case SW-2 and for the limiting cases ($S/\overline{\delta} \gg 1$ and $\ll 1$) and larger than 1 for the intermediate cases ($S/\overline{\delta} \approx 1$). For the limiting cases $S/\overline{\delta} \gg 1$ ($S/\overline{\delta} = 2.28$ and 3.63), the magnitude of $\hat{y}$ measured at $z_{sheet}/\overline{\delta} \approx 0.5$ is closer to that of the reference smooth-wall case SW-2 at $z_{sheet}/\delta_s = 0.46$ (dashed line in figure~\ref{fig:yhat}b), while in the limit $S/\overline{\delta} \ll 1$, $\hat{y}$ measured at $z_{sheet}/\overline{\delta} = 0.18$ is approximately equal to that of SW-2 at $z_{sheet}/\delta_s = 0.24$ (dash-dot line in figure~\ref{fig:yhat}b). This suggests that in the limiting cases where $S \gg \overline{\delta}$ or $S \ll \overline{\delta}$, the meandering of the secondary flows reverts to smooth-like behaviour. 

It is also important to note that while the contours in figure~\ref{fig:condave_eq1} show the conditionally averaged (and filtered) fluctuating velocity field $u'$, the condition points (`+' in figure~\ref{fig:spines}) are extracted from $\widetilde{u}'$ for SR cases (which contains the stationary component of the secondary flows $\widetilde{U}$) and from $u'$ for the reference smooth-wall case SW-2 ($\widetilde{U} = 0$ for this case). As a direct comparison, we compute the same conditional average of the maxima and minima of the low-speed structures with the spines extracted from $u'$ (the non-stationary component of secondary flow) instead of $\widetilde{u}'$ for all SR cases. An example of this analysis is given in appendix~\ref{app} for case SR50 ($S/\overline{\delta} = 0.62$). The magnitude of $\hat{y}$ extracted from the $u'$ field (`\textcolor{red}{$\times$}' symbol in figure~\ref{fig:yhat}b) is approximately equal to that extracted from $\widetilde{u}'$ for the limiting cases $S/\overline{\delta} \gg 1$ and $\ll 1$, and slightly smaller for the intermediate cases $S/\overline{\delta} \approx 1$. However, in general this illustrates that the choice of $u'$ or $\widetilde{u}'$ for the condition vector makes very little difference in terms of the salient trends exhibited in figure~\ref{fig:yhat}(b). Regardless of the condition vector, the meandering amplitude of the secondary flows exhibit a clear peak when $S/\overline{\delta} \approx 1$. 

\section{Secondary flows and large-scale structures}

\begin{figure}
	\centering
	\includegraphics[height=10cm, width=10cm, keepaspectratio]{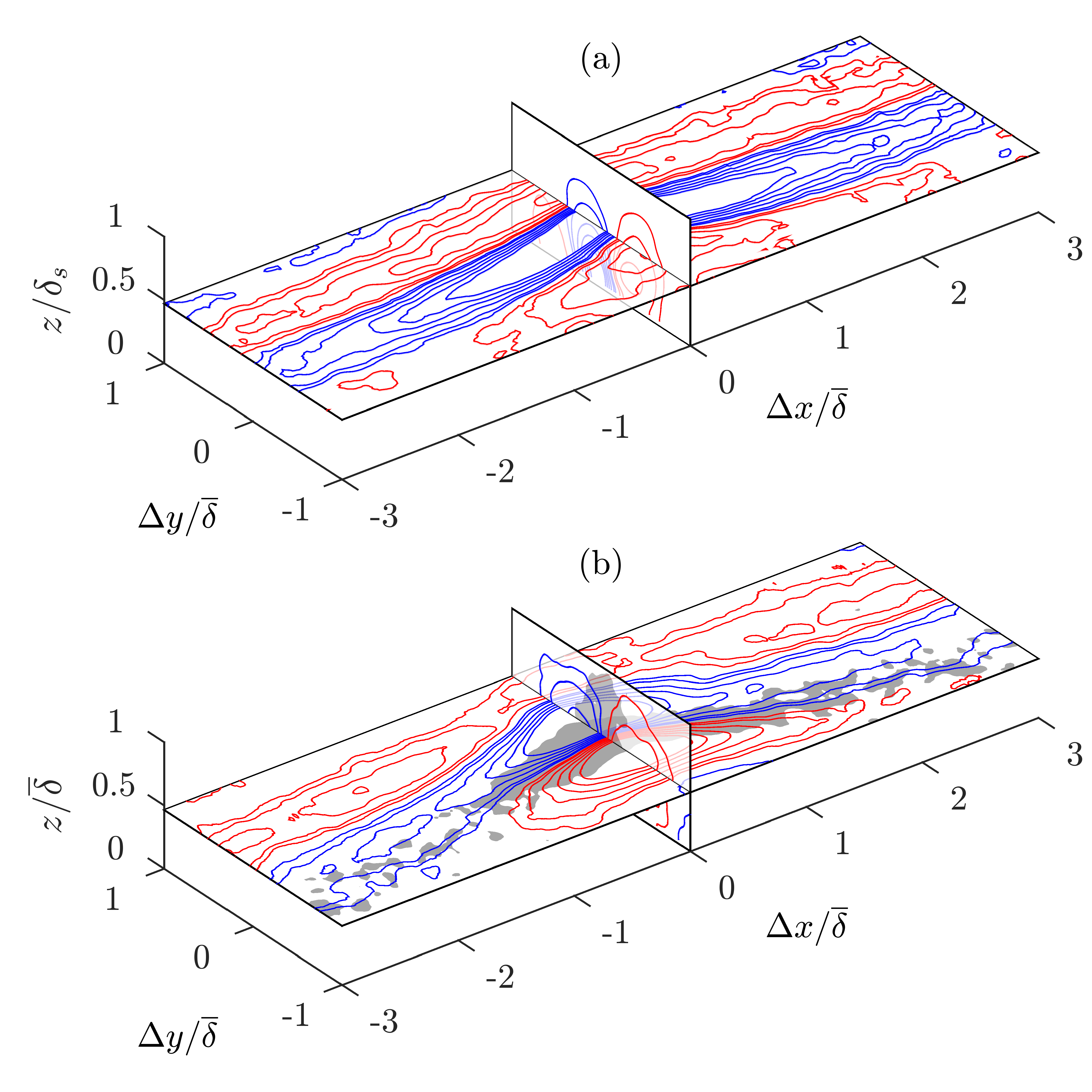}
	\vspace{0cm}
	\caption{Illustrations of meandering low-speed structures flanked by two high-speed structures constructed by conditionally averaged $u'$ from SPIV and WPPIV plane for (a) the reference case SW-2 and (b) case SR100 ($S/\overline{\delta} = 1.35$, structures extracted from the red region in figure~\ref{fig:spines}d), conditioned at the minima of the detected low-speed structures illustrated in figure~\ref{fig:spines}. Solid red lines (\solidline[red]) show $u_f'U_{\infty} =  0.05, 0.1, 0.2, ..., 0.5$, blue solid lines (\solidline[blue]): $-0.5, -0.4, ..., -0.1, -0.05$. In (b), gray-shaded contours indicate vorticity of the conditionally averaged velocity field $\omega'_x \overline{\delta}/U_{\infty} = 0.2$ and $\omega'_z \overline{\delta}/U_{\infty} = 0.8$ for case SR100 in $y$--$z$ and $x$--$y$ plane, respectively.}
	\label{fig:condave_3d}
\end{figure}

The similarities and differences between secondary flows (intermediate cases, figure~\ref{fig:condave_eq1}d,e) and the large-scale motions (SW case, figure~\ref{fig:condave_eq1}a) raise many important questions. Firstly, is it possible that secondary flows due to spanwise heterogeneity and large-scale motions share similar formation mechanisms? Some parallels regarding the formation mechanism have been suggested. For example, \cite{leejh2019}, when studying VLSMs in canonical smooth wall pipe flow, have associated their formation with instantaneous lateral variations of the wall shear stress of homogeneous turbulent pipe flows. Secondary flows, on the other hand are also associated with stationary or persistent lateral variations in the wall shear stress \citep{townsend1976}. In this sense, both naturally occurring large-scale structures and secondary flows are both triggered by lateral stress variations (the former convecting and transient, and the latter stationary). Figure~\ref{fig:condave_3d}(a,b) show the 3-D reconstruction of the large-scale streaks from $xy$- and $yz$-plane for case SW-2 and SR100 ($S/\overline{\delta} = 1.35$), respectively\footnote{It should be noted that the condition vectors for the ensemble averaging differ between the $xy$- and $yz$-plane and these figures are only intended to show the representation of the high- and low-speed structures above the reference smooth wall case and an intermediate case.}. The yawed and inclined asymmetric roll modes can be inferred from the in-plane vorticity (grey-shaded contours in figure~\ref{fig:condave_3d}b) calculated from the conditionally averaged velocity fields: $\omega_x' \equiv \partial w_f'/\partial y - \partial v_f'/\partial z$ in $y$--$z$ plane and $\omega_z' \equiv \partial v_f'/\partial x - \partial u_f'/\partial y$ in $x$--$y$ plane. A cursory look at figure~\ref{fig:condave_3d} reveals that the meandering is more prominent in the intermediate cases where $S/\overline{\delta} \approx 1$ (secondary flows, figure~\ref{fig:condave_3d}b) than that of the reference smooth-wall case (LSM/VLSM, figure~\ref{fig:condave_3d}a). However, both structures show similarities to the streak-vortex models for the naturally occurring large-scale structures proposed in \cite{jeong1997,waleffe2001,schoppa2002,flores2010,cossu2017,degiovanetti2017}. These asymmetric streak-vortex structures have also been elucidated in the log region from both PIV of converging-diverging (C-D) riblets, which also induce secondary flows, and in direct numerical simulations (DNS) of smooth-wall channel flow \citep{kevin2019b}. This leads to other questions: if it is possible that both secondary flows and large-scale structures share a similar formation mechanism, is it also possible that the secondary flows are just phase-locked large-scale structures? If that is the case, why do the secondary flows strongly meander only when $S/\overline{\delta} \approx 1$, as shown in figure~\ref{fig:yhat}(b)? Finally, we ask if the secondary flows and the large-scale structures coexist with each other in certain heterogeneous wavelengths. In the next section of this study, we attempt to address these questions.

\subsection{Phase locking}
\label{sub:spectra}

The 1-D spanwise energy spectra, related to spanwise wavenumber $k_y$ and wavelength $\lambda_y$, are computed from the fluctuating velocity field obtained from SPIV measurements. At any wall-normal location $z$ in the FOV, the non-normalised two-point correlation of the fluctuating velocity component $u'$ is given by
\begin{equation}
\hat{R}_{u'u'} (\Updelta y, z) = \overline{u'(y,z)u'(y+\Updelta y,z)}
\label{eq:Rab_fixz_slidey_nonorm}
\end{equation} 
where the overbar denotes ensemble averaging at all $y$ and for all SPIV realizations and $\Updelta y$ is the spanwise shift. The Fourier transformation of $\hat{R}_{u'u'}$ yields the energy spectra of the fluctuating velocity component $u'$
\begin{equation}
\Upphi_{u'u'} = \int_{-\infty}^{\infty} \hat{R}_{u'u'} \; \mathrm{e}^{-j 2\pi k_y \Updelta y} \; \mathrm{d}(\Updelta y)
\label{eq:Phiab_dy}
\end{equation} 
where $k_y$ is the spanwise wavenumber and $\lambda_y = 2\pi/k_y$ is the spanwise wavelength. The relationship between the Reynolds stress component $\overline{u'u'}$ and $\Upphi_{u'u'}$ is given by  
\begin{equation}
\overline{u'u'} = \int_{0}^{\infty} \Upphi_{u'u'} \; \mathrm{d}(k_y)
\label{eq:Phiab_dy_int}
\end{equation}
Due to the limited spanwise extent of the FOV, the velocity field is padded with zeros in $y$ to increase the length of the signal to three times the total width of the FOV (from $4\delta_s$ to about $12\delta_s$) and thus increase the resolution in $k_y$ and $\lambda_y$. This method for 1-D spanwise energy spectra computation, however, must be applied with caution for the spanwise heterogeneous roughness cases. Figure~\ref{fig:U_decomposed}(f) shows the instantaneous turbulent fluctuation component $u'$ for case SR50 ($S/\overline{\delta} = 0.62$), whose mean (for all snapshots) across the repeating period of the spanwise heterogeneous roughness is zero. Its variance $\overline{u'u'}$, however, is heterogeneous in $y$ (see figure~\ref{fig:U_decomposed}c). Using this method, integration of the energy spectra across $k_y$ in (\ref{eq:Phiab_dy_int}) results in the \emph{average} of $\overline{u'u'}$ in $y$, and yields no information regarding the spanwise heterogeneity of the flow. 

\begin{figure}
	\centering
	\includegraphics[width=13cm, height=8.1cm, keepaspectratio]{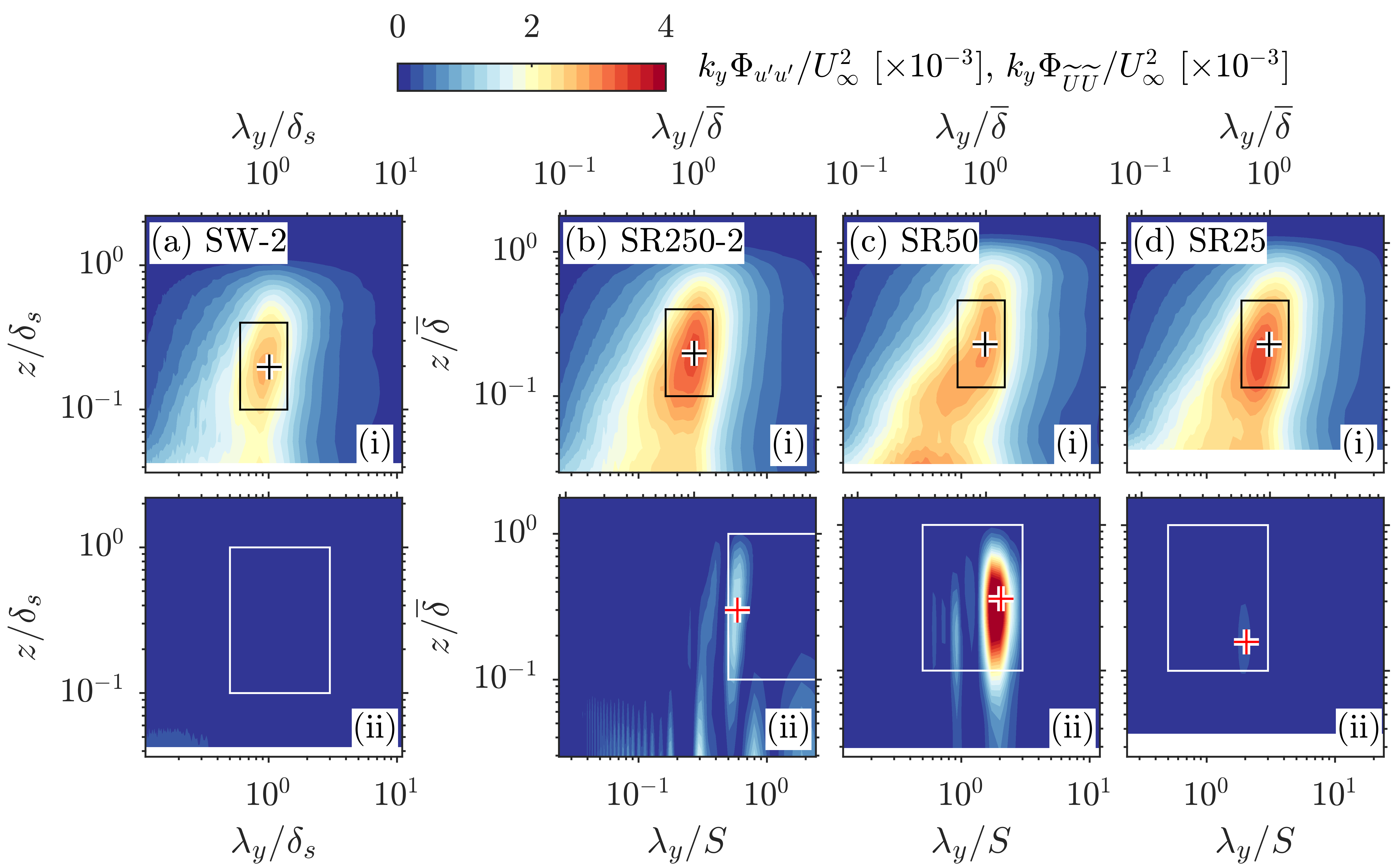}
	
	\caption{Contours of premultiplied energy spectra (i) $k_y \Upphi_{u'u'}$ and (ii) $k_y \Upphi_{\widetilde{U}\widetilde{U}}$ as functions of $\lambda_y$ and $z$ for (a) reference case SW-2 and SR cases: (b) SR250-2 ($S/\overline{\delta} = 3.63$), (c) SR50 ($S/\overline{\delta} = 0.62$), (d) SR25 ($S/\overline{\delta} = 0.32$). Data are obtained from SPIV measurements. In (i), `+' marks the peak of energy spectra in the outer layer at $z/\overline{\delta} = 0.2$ and $\lambda_y/\overline{\delta} = 1$ (b--d), $z/\delta_s = 0.2$ and $\lambda_y/\delta_s = 1$ in (a). Black boxes are the area of integration in~(\ref{eq:spiv_spectra_int}) In (ii), `\textcolor{red}{+}' marks the spanwise wavelengths related to the mean secondary flows. White boxes are the area of integration in~(\ref{eq:spiv_spectra_int_tilde}) and (\ref{eq:spiv_spectra_int_tilde_sw}).}
	\label{fig:spiv_spectra}
\end{figure}

Figure~\ref{fig:spiv_spectra}(i) shows the premultiplied energy spectra $k_y \Upphi_{u'u'}$ for the reference smooth wall case SW-2 (figure~\ref{fig:spiv_spectra}a(i)) and SR cases: SR250-2, SR50, and SR25 (figure~\ref{fig:spiv_spectra}b(i)--d(i), respectively). In general, all SR cases have higher energy than the reference smooth wall case SW-2, which is expected since the presence of surface roughness results in higher magnitude of energy (when normalised by $U_{\infty}$). Further, all cases (SR and SW-2) also show an outer peak (`+') at $z/\overline{\delta} \approx 0.2$ and $\lambda_y/\overline{\delta} \approx 1$ ($z/\delta_s \approx 0.2$ and $\lambda_y/\delta_s \approx 1$ for SW-2) which is associated with the width and spacing between large-scale low- and high-speed streaks for the large-scale and very large-scale motions (LSM/VLSM). It is important to note that $k_y \Upphi_{u'u'}$ does not contain information regarding the stationary (time-averaged) secondary flows. Instead, this is contained in $\widetilde{U}$ (figure~\ref{fig:spiv_spectra}(ii)). Figure~\ref{fig:spiv_spectra}(i) contains information about non-stationary turbulent fluctuations, and these seem largely similar between the smooth and heterogeneously rough surfaces. The contours of spanwise energy spectra for SR cases calculated from the SPIV data also do not contain a clear imprint of the time-dependant (meandering) secondary flows, unlike the 1-D streamwise energy spectra $k_x \Upphi_{u'u'}$ \citep[see][figure 13 and 14]{wangsawijaya2020}, which is expected since this assumed meandering behaviour is related primarily to the phase of the spanwise Fourier modes which are not analysed in the amplitude spectra (although increased meandering might affect the magnitude of the resolved $k_y \Upphi_{u'u'}$ energy).

It is also worth noting that compared to other SR cases, the magnitude of $k_y \Upphi_{u'u'}$ is the lowest for case SR50 (figure~\ref{fig:spiv_spectra}c(i)) -- the case which shows the strongest outer peak in $k_x \Upphi_{u'u'}$ and the most prominent meandering \citep{wangsawijaya2020}. To understand this phenomenon, another attempt is made to examine the characteristics of the secondary flows that appear in the spanwise energy spectra. Instead of the turbulent fluctuations, the energy spectrograms are computed for the time-averaged spatial velocity fluctuations $\widetilde{U}$ (figure~\ref{fig:U_decomposed}e), which carries the footprint of the stationary (time-averaged) secondary flows. The method for computation is similar to that of turbulent fluctuations in (\ref{eq:Rab_fixz_slidey_nonorm})--(\ref{eq:Phiab_dy_int}). To ensure zero mean across $y$, the FOV is reflected for the largest $S$ cases (SR250-2 and SR160) so that it captures one spanwise roughness wavelength $\Lambda = 2S$. For the cases where $S/\overline{\delta} \lesssim 1$: SR100, SR50, and SR25, the FOV is clipped so that it captures an integer number of spanwise wavelengths, $\Lambda$, 2$\Lambda$, and 4$\Lambda$, respectively.  

Figure~\ref{fig:spiv_spectra}(ii) shows the premultiplied energy spectra of $\widetilde{U}$. As expected, the magnitude of $k_y \Upphi_{\widetilde{U}\widetilde{U}}$ is zero for the reference smooth wall case SW-2 in figure~\ref{fig:spiv_spectra}(a)(ii). Clear, prominent modes in the energy spectra are observed in all SR cases in figure~\ref{fig:spiv_spectra}(b)(ii)--(d)(ii). These peaks are related to the mean secondary flows and are expected to occur at $\lambda_y = 2S$, i.e. the spanwise roughness wavelength, except for the case where $S/\overline{\delta} > 1$. Here, for case SR250-2 ($S/\overline{\delta} = 3.63$, figure~\ref{fig:spiv_spectra}b(ii)), the peak is observed at $\lambda_y/S \approx 0.6$ ($\lambda_y/\overline{\delta} \approx 2$). This observation is in line with the finding in \cite{wangsawijaya2020}; that the secondary flow size is capped by $\overline{\delta}$ instead of $S$ for this case (see figure~\ref{fig:limitcase}a). The prominent mode is locked at $\lambda_y = 2S$ for cases where $S/\overline{\delta} \leq 1$. Contours of $k_y \Upphi_{\widetilde{U}\widetilde{U}}$ show a very strong mode at $\lambda_y/S = 2$ for case SR50 ($S/\overline{\delta} = 0.62$, figure~\ref{fig:spiv_spectra}c(ii)) and a much weaker one for case SR25 ($S/\overline{\delta} = 0.32$, figure~\ref{fig:spiv_spectra}d(ii)). Similar to previous observations regarding the strength of mean secondary flows \citep{wangsawijaya2020}, the strongest magnitude of energy in $\widetilde{U}$ occurs at SR50 ($S/\overline{\delta} = 0.62$) and it decreases as $S$ approaches $S/\overline{\delta} \ll 1$.   

The premultiplied energy spectra of $u'$ and $\widetilde{U}$ seem to show opposite tendency for the SR50 case: when compared to the rest of the SR cases, the magnitude of $k_y \Upphi_{u'u'}$ exhibits a minimum while the magnitude of $k_y \Upphi_{\widetilde{U} \widetilde{U}}$ is maximum (see figure~\ref{fig:spiv_spectra}c(i) and~\ref{fig:spiv_spectra}c(ii)). To further examine this, $k_y \Upphi_{u'u'}$ is integrated across the area enclosing the outer peak (black boxes in figure~\ref{fig:spiv_spectra}(i)): $0.6 \leq \lambda_y/\overline{\delta} \leq 1.4$ and $0.1 \leq z/\overline{\delta} \leq 0.4$,
\begin{equation}
I_{\Phi_{u'u'}} = \frac{1}{0.3\overline{\delta}} \int\limits_{0.1\overline{\delta}}^{0.4\overline{\delta}}  \int\limits_{10\pi/3\overline{\delta}}^{10\pi/7\overline{\delta}} \frac{\Upphi_{u'u'}}{U_{\infty}^2} \mathrm{d}k_y \; \mathrm{d}z
\label{eq:spiv_spectra_int}
\end{equation} 
For the reference smooth wall cases SW-2, $\delta_s$ is used instead of $\overline{\delta}$. $k_y \Upphi_{\widetilde{U} \widetilde{U}}$ for SR cases is integrated across the area enclosing the prominent $\widetilde{U}$ modes (white boxes in figure~\ref{fig:spiv_spectra}b(ii)-d(ii)): $0.5 \leq \lambda_y/S \leq 3$ and $0.1 \leq z/\overline{\delta} \leq 1$,
\begin{equation}
I_{\Phi_{\widetilde{U} \widetilde{U}}} = \frac{1}{0.9\overline{\delta}} \int\limits_{0.1\overline{\delta}}^{\overline{\delta}}  \int\limits_{4\pi/S}^{2\pi/3S} \frac{\Upphi_{\widetilde{U} \widetilde{U}}}{U_{\infty}^2} \mathrm{d}k_y \; \mathrm{d}z
\label{eq:spiv_spectra_int_tilde}
\end{equation}
For the reference smooth wall case in figure~\ref{fig:spiv_spectra}(a)(ii), the integration area encompasses $0.5 \leq \lambda_y/\delta_s \leq 3$ and $0.1 \leq z/\delta_s \leq 1$, 
\begin{equation}
I_{\Phi_{\widetilde{U} \widetilde{U}}} = \frac{1}{0.9\delta_s} \int\limits_{0.1\delta_s}^{\delta_s}  \int\limits_{4\pi/\delta_s}^{2\pi/3\delta_s} \frac{\Upphi_{\widetilde{U} \widetilde{U}}}{U_{\infty}^2} \mathrm{d}k_y \; \mathrm{d}z
\label{eq:spiv_spectra_int_tilde_sw}
\end{equation}

\begin{figure}
	\centering
	\includegraphics[width=6.5cm, height=4cm, keepaspectratio]{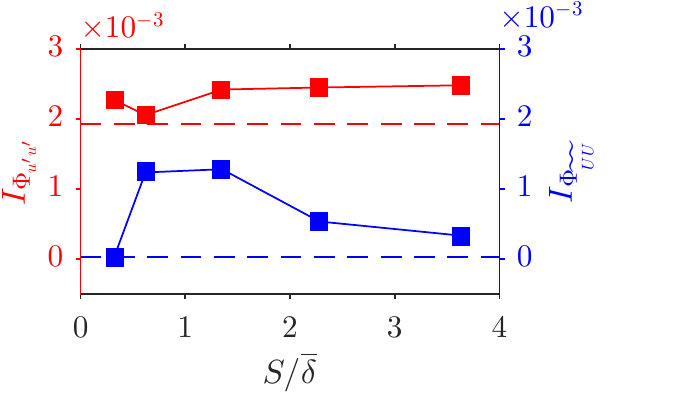}
	
	\caption{Integrated energy spectra of $u'$ ($I_{\Phi_{u'u'}}$, \squareline[red]) and $\widetilde{U}$ ($I_{\Phi_{\widetilde{U}\widetilde{U}}}$, \squareline[blue]) as a function of $S/\overline{\delta}$ for SR cases. Red and blue dashed lines (\dashedline[red] and \dashedline[blue]) are $I_{\Phi_{u'u'}}$ and $I_{\Phi_{\widetilde{U}\widetilde{U}}}$ for the reference smooth wall SW-2, respectively.}
	\label{fig:spiv_spectra_int}
\end{figure}

Integrated energy spectra of $u'$ and $\widetilde{U}$ are shown in figure~\ref{fig:spiv_spectra_int} as functions of $S/\overline{\delta}$ (dashed lines show the reference smooth wall case SW-2). The lines in figure~\ref{fig:spiv_spectra_int} suggest a relationship between $u'$ and $\widetilde{U}$: as $S/\overline{\delta}$ approaches 1, $\widetilde{U}$ and $u'$ reach a maxima and minima, respectively. As $S/\overline{\delta} \gg 1$, the rise of $u'$ is accompanied by a decrease in $\widetilde{U}$. In some sense this fits with the conceptual picture that the secondary flows are spanwise-locked large-scale structures. Under this assumption, SR50 ($S/\overline{\delta} = 0.62$) presumably is the case where the locking is most complete, since the secondary flows are the strongest and fill the entire wall-normal and spanwise extent of the boundary layer \citep{wangsawijaya2020}. If the secondary flows can really be viewed as locked turbulent structures, it seems reasonable to assume that for heterogeneous surfaces that generate secondary flows some of the large-scale structure energy will be transferred from turbulent fluctuations $u'$ to the dispersive component $\widetilde{U}$. This interplay between turbulent fluctuations and the dispersive components (as one goes up, the other one goes down and vice versa) is strongly reminiscent of similar observations by \cite{nikora2019} of secondary flows in rough-bed open channel flows and by \cite{modesti2018} of secondary flows in a square duct. In the latter case, \cite{modesti2018} also observed that when the secondary flows are artificially suppressed, the turbulent fluctuations rise to compensate the loss of dispersive component. Taken together, a possible interpretation of these results would be that the $\widetilde{U}$ component may be an indicator of the efficacy of the surface at locking turbulent structures in place. For example, in the smooth surface, $\widetilde{U}$ is zero (blue dashed line in figure~\ref{fig:spiv_spectra_int} and figure~\ref{fig:spiv_spectra}a(ii)), and the turbulent structures are randomly occurring  (completely unlocked), so all energy ends up in the $u'$ component. It is noted that the cases where $S/\overline{\delta} \ll 1$ and $S/\overline{\delta} \gg 1$ in figure~\ref{fig:spiv_spectra_int} (SR25, $S/\overline{\delta} = 0.32$ and SR250-2, $S/\overline{\delta} = 3.63$) approach this condition, with the integrated energy in $\widetilde{U}$ approaching the smooth wall reference case. A possible conclusion from the above analysis is that for $S/\overline{\delta} \approx 1$, the turbulent structures are optimally locked or trapped by the surface, and hence energy from $u'$ is surrendered into $\widetilde{U}$ (which respectively reach a minima and a maxima at these $S/\overline{\delta}$). However, this suggested behaviour is complicated somewhat by the observed meandering behaviour in \S\ref{sub:sond_eq1}. It is noted in this study, and in that of \cite{wangsawijaya2020}, that the meandering of the secondary flows increases above that of the naturally occurring turbulent structures when $S/\overline{\delta} \rightarrow 1$. Such behaviour is not easily reconcilable with the notion of secondary flows as locked turbulent structures and is perhaps indicative of some additional instability or forcing when secondary flows are locked at particular spanwise wavelengths. 

\subsection{Coexistence between secondary flows and the large-scale structures}

\begin{figure}
	\centering
	\includegraphics[width=13cm, height=4.2cm, keepaspectratio]{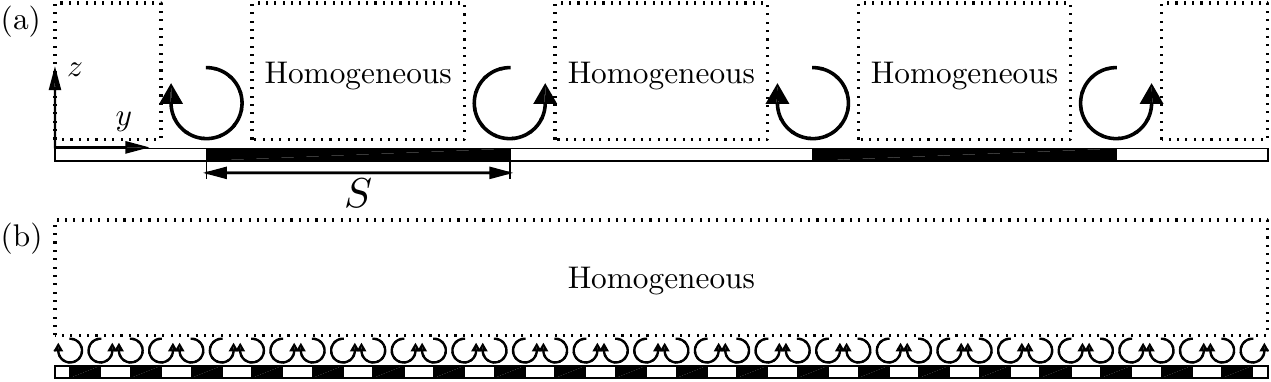}
	\caption{Illustration of the limiting cases in spanwise heterogeneous roughness: (a) $S/\overline{\delta} \gg 1$ and (b) $S/\overline{\delta} \ll 1$. Areas bounded by \dottedline[black] show regions approaching spanwise homogeneity.}
	\label{fig:limitcase}
\end{figure}

Coexistence between the VLSM and secondary flows was hinted at in \cite{zampiron2020} at larger spanwise wavelengths, where these two features exist at a very different scale in the energy spectrograms. At both limits, when $S/\overline{\delta} \gg 1$ and when $S/\overline{\delta} \ll 1$, the secondary flows are confined in certain areas of the surface such that we might expect the LSM/VLSM to exist in a relatively unaltered form for large areas of the flow, especially when we are far from the secondary flows, e.g. far from the roughness transition for the case with $S \gg \overline{\delta}$, or $z > S$ for the case where $S \ll \overline{\delta}$ (figure~\ref{fig:limitcase}). What follows is an attempt to separate the secondary flows due to spanwise heterogeneity and (presumably) unaltered, naturally occurring LSM/VLSM in the limiting cases: $S/\overline{\delta} \gg 1$ (\S\ref{sub:sond_gg1}) and $S/\overline{\delta} \ll 1$ (\S\ref{sub:sond_ll1}). Conditional averaging and POD analysis are conducted for the two largest wavelength cases: SR250-2 ($S/\overline{\delta} = 3.63$) and SR160 ($S/\overline{\delta} = 2.28$), and the smallest wavelength, SR25 ($S/\overline{\delta} = 0.32$). 

\subsubsection{Limiting case: $S/\overline{\delta} \gg 1$}
\label{sub:sond_gg1}

Figure~\ref{fig:spines}(b,c) show the detected low-speed structures and the minima of the spines of these structures at an instantaneous velocity field of limiting cases SR250-2 ($S/\overline{\delta} = 3.63$) and SR160 ($S/\overline{\delta} = 2.28$), respectively, taken from WPPIV measurements approximately at the center of the mean secondary flows ($z/\overline{\delta} \approx 0.5$). Instantaneously, the large-scale structures induced by spanwise heterogeneity meander about a certain spanwise location, which results in $\delta$-scaled secondary flows in the time-averaged velocity field near the interface between rough and smooth strips \citep[see][figure 11]{wangsawijaya2020}. Here, it is assumed that the low-speed structures related to the secondary flows (upwelling motions) meander about the spanwise location of common flow up $y_u \pm l_y$, where $l_y$ is the spanwise extent of the mean secondary flows. This area is shaded red in figure~\ref{fig:spines}. Far removed from the secondary flows, the flow becomes locally homogeneous (either homogeneously smooth or homogeneously rough). In figure~\ref{fig:spines}(b), areas shaded in blue (closer to the centerline of smooth and rough strips) represent this condition. It is assumed that secondary flows occur in the red-shaded area, while LSM/VLSM occur in the blue-shaded area (these two terms are used here to distinguish between structures that are imposed or locked by the roughness heterogeneity and naturally occurring turbulent structures, respectively). Histograms in figure~\ref{fig:hist_yref}(b,c) show that the secondary flows (red bars) comprise 41\% and 74\% of all low-speed structures detected across the FOV for case SR250-2 and SR160, respectively (compared to respective areas occupied by the red regions of 37\% and 54\% of the FOV), and thus LSM/VLSM (blue bars) comprise the rest (59\% and 26\%). 

\begin{figure}
	\centering
	\includegraphics[height=17.1cm, width=13cm, keepaspectratio]{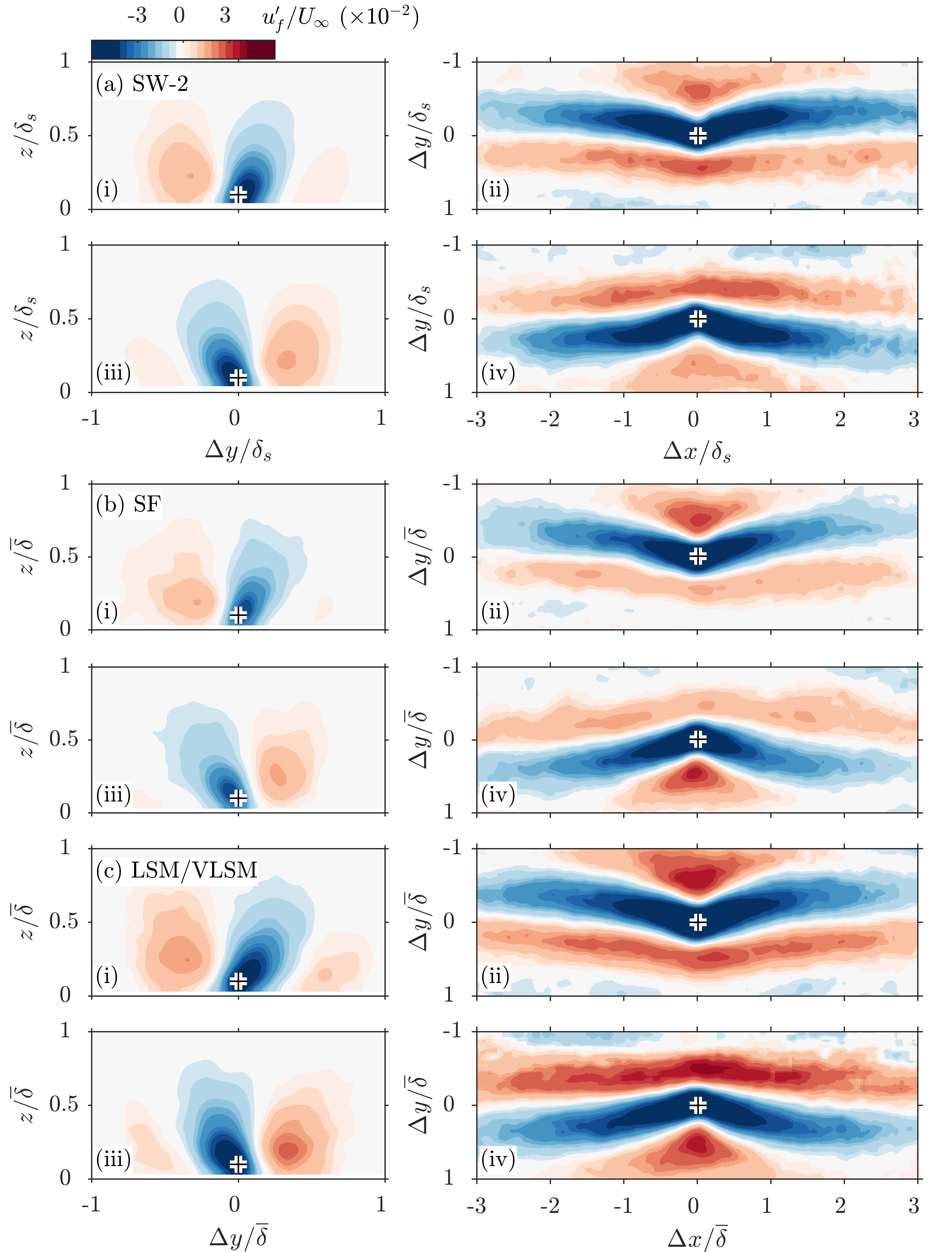}
	\caption{Contours of conditionally averaged turbulent fluctuation $u'$ for (a) reference case SW-2 at $z/\delta_s = 0.46$ and (b,c) SR250-2: conditioned at low-speed structures associated with (b) secondary flows (SF) and (c) LSMs/VLSMs. Conditions: (i) $\widetilde{u}' < 0$ and $v' > 0$, (iii) $\widetilde{u}' < 0$ and $v' < 0$, (ii) $y$-minima and (iv) $y$-maxima of detected spine-fitted low-speed structures ($\widetilde{u}'/U_{\infty} < -0.03$) illustrated in figure~\ref{fig:spines}(b).}
	\label{fig:condave_sr250}
\end{figure}

Similar to the analysis conducted in \S\ref{sub:sond_eq1}, the conditional average is also calculated at the $y$-minima and $y$-maxima of the spines fitted to detected low-speed structures and in the $yz$-plane from SPIV measurements, using a different but related condition vector. For the $yz$-plane, the average is computed at $z_{ref}/\overline{\delta} = 0.1$, based on the condition of $\widetilde{u}'$ and $v'$ (see \S\ref{sub:sond_eq1}). Figure~\ref{fig:condave_sr250} shows the conditional average of case SR250-2 (figure~\ref{fig:condave_sr250}b,c) compared to the reference smooth wall case SW-2 (figure~\ref{fig:condave_sr250}a). The conditional average for the secondary flows (`SF') due to spanwise heterogeneity given in figure~\ref{fig:condave_sr250}(b) is computed at the spanwise location of common flow up (\circled[u]) in subplot (i) and (iii). It is computed at the minima and maxima of the detected low-speed structures inside the red-shaded area (figure~\ref{fig:spines}b) in subplot (ii) and (iv). The conditional average for the LSM/VLSM is computed at $y_{ref}/S = -0.72$, close to the centerline of a rough strip ($y/S = -1$). A one-sided leaning tendency is observed in the conditionally averaged secondary flows (figure~\ref{fig:condave_sr250}b). However, both secondary flows and LSM/VLSM (figure~\ref{fig:condave_sr250}c) have similar shape and size (when scaled by $\overline{\delta}$) to the LSM/VLSM in the reference smooth wall case.

A cursory inspection of figures~\ref{fig:condave_sr250}(b) and (c) suggest that the LSM/VLSM are stronger (higher in magnitude) than the secondary flows. This is expected since the secondary flows are assumed to occur above the interface between rough and smooth strips, while the detected LSMs/VLSMs are mostly obtained above rough strips (figure~\ref{fig:swirl}a), where the turbulent energy is higher (note that the conditionally averaged $u_f'$ in figure~\ref{fig:condave_sr250} is normalised by $U_{\infty}$). This observation, although not shown for brevity, also holds for SR160 ($S/\overline{\delta} = 2.28$). Comparing figures~\ref{fig:condave_sr250}(a) and (b), we observe that the LSM/VLSM of the reference smooth-wall case SW-2 is slightly stronger than the secondary flows of SR250-2 case. However, despite these minor differences, the overarching impression from figure~\ref{fig:condave_sr250} is the striking similarity between the smooth wall LSMs/VLSMs and both the secondary flows and LSMs/VLSMs that occur in the large scale limiting spanwise heterogeneity case ($S \gg \overline{\delta}$). 

A comparison of the magnitude of meandering between the secondary flows and LSM/VLSM is shown by the blue and red symbols in figure~\ref{fig:yhat}(b). Recall that this meandering amplitude is measured from the `spines' extracted from the conditionally averaged low-speed structures (figure~\ref{fig:condave_eq1}a). This comparison is only possible for the range of spanwise heterogeneous wavelengths  ($1.35 \leq S/\overline{\delta} \leq 3.63$), where the red and blue regions of figure~\ref{fig:spines} can both be defined. At smaller wavelengths, the secondary flow regions (defined as $y_u \pm l_y$) occupy the entire spanwise domain and so the LSM/VLSM region cannot be identified. The trend shows that $\hat{y}$ (normalised by the $z_{sheet}$/size of the roll modes) of secondary flows increases as $S/\overline{\delta}$ approaches 1, but $\hat{y}$ of the LSM/VLSM, whether it is extracted from $\widetilde{u}'$ or $u'$ fields, remains approximately constant. More importantly, as $S/\overline{\delta}$ becomes large in case SR250-2, we note from figure~\ref{fig:yhat}(b) that the meandering amplitude of both the secondary flows and the LSMs/VLSMs seem to converge to the same value, which is close to that recorded for the smooth wall. Again this reconfirms the general underlying similarity between the conditional features presented for the smooth wall and case SR250-2 in figure~\ref{fig:condave_sr250}.

\subsubsection{Limiting case: $S/\overline{\delta} \ll 1$}
\label{sub:sond_ll1}

For the smallest $S/\overline{\delta}$ case, SR25 ($S/\overline{\delta} = 0.32$), the WPPIV measurement plane is set as close as possible to the centre of the secondary flows at $z/\overline{\delta} = 0.18$, while the diameter of the secondary flows is approximately $0.32\overline{\delta}$. As shown in figure~\ref{fig:swirl}(c), the secondary flows fill the entire spanwise extent at this wall height (see also red shaded area in figure~\ref{fig:spines}f). A similar conditional average at the minima and maxima of the detected low-speed structures in the $x$--$y$ plane and at the common flow up in the $y$--$z$ plane as that in \S\ref{sub:sond_eq1} and \S\ref{sub:sond_gg1} is also computed for SR25 case. Figure~\ref{fig:condave_eq1}(f) shows that the conditionally averaged low-speed structures for this case do not seem to show a marked difference to the LSM/VLSM in the reference smooth wall case. The meandering amplitude for this case (figure~\ref{fig:yhat}b) is also approximately equal to that of the reference smooth wall case at $z/\delta_s = 0.24$.

These observations for the $S/\overline{\delta} \ll 1$ case reveal the problem of using spanwise locations to separate secondary flows and LSM/VLSM. This strategy works for $S/\overline{\delta} \gg 1$ cases because both secondary flows and LSM/VLSM are both $\delta$-scaled and the secondary flows are confined about the roughness interface, while the rest of the flow approaches local homogeneity (smooth wall or homogeneous rough wall). However, for the other limit where $S/\overline{\delta} \ll 1$, the secondary flows and LSM/VLSM coexist in different scales. The LSM/VLSM are $\delta$-scaled, while the secondary flows are scaled by $S$ ($S \ll \delta$) and aligned according to the roughness strips. For case SR25, specifically, the spanwise scale of the secondary flows is approximately $0.32\overline{\delta} \approx S$. Hence, to study the coexistence between LSM/VLSM and secondary flows in this limit where $S/\overline{\delta} \ll 1$, a different method is necessary to separate the two.   

We propose `snapshot' proper orthogonal decomposition (POD) as a possible method to separate LSM/VLSM and secondary flows in the SR25 case. POD for fluid mechanics applications was first introduced by \cite{lumley1967} and the algorithm used in the present study is the so-called `snapshot' POD (\citealp{sirovich1987}, see also \citealp{meyer2007} for details). In the context of PIV data, each instantaneous velocity field is considered as a `snapshot' and the fluctuating velocity components of snapshot $n$, $\boldsymbol{u'}^{n} = (u'^{n}, v'^{n}, w'^{n})$, can be written as a linear expansion
\begin{equation}
\boldsymbol{u'}^{n} = \sum_{i=1}^{N} \boldsymbol{a}_{i}^{n} \boldsymbol{\phi}^{i}
\label{eq:pod}
\end{equation}   
where $\boldsymbol{\phi}^{i}$ is the $i$th POD mode, $\boldsymbol{a}_{i}^{n}$ is the POD coefficient of mode $i$ of snapshot $n$, and $N$ is the total number of snapshots. In snapshot POD, the number of resolved POD modes is equal to the number of snapshots, which corresponds to 4800 for the spanwise heterogeneous cases in the $yz$-plane (1200 for the reference smooth wall case SW-2) and 1200 snapshots in the $xy$-plane (600 for SW-2). It is noted that the POD modes $\boldsymbol{\phi}^{i}$ are ordered according to their contribution to the total turbulent kinetic energy such that the first mode $\phi^{1}$ has the largest fraction of total energy, followed by the second mode $\phi^{2}$, and so forth. 

\begin{figure}
	\centering
	\includegraphics[height=7cm, width=13cm, keepaspectratio]{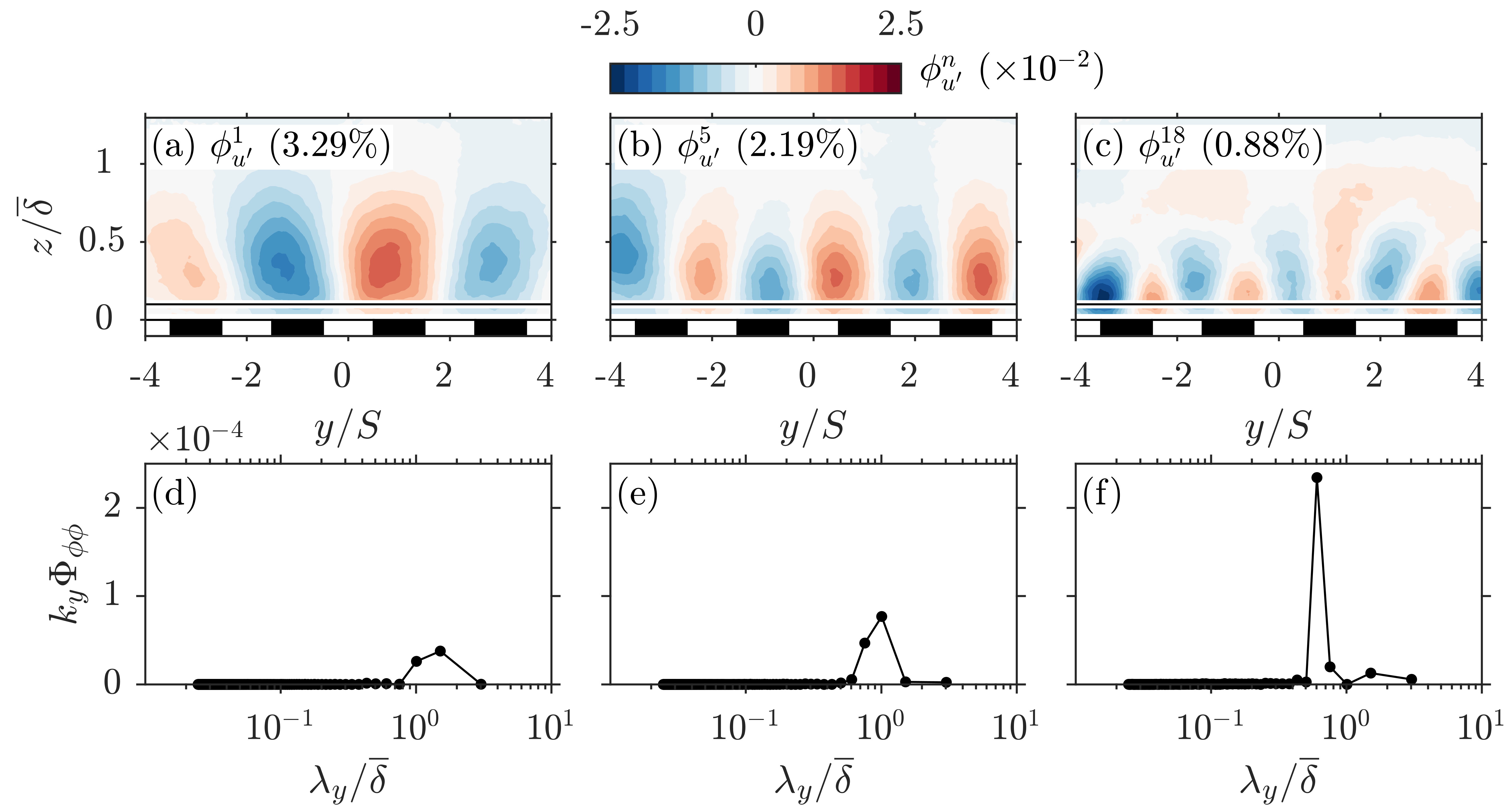}
	\caption{POD modes of $u'$ for SR25 case in $yz$-plane: (a,d) mode 1, (b,e) mode 5, and (c,f) mode 18. Brackets on top left of the figures show the fraction of energy of the POD modes. (d,e,f) are 1-D premultiplied energy spectra of the modes $k_y \Upphi_{\phi \phi}$ as a function of $\lambda_y$, computed at $z/\overline{\delta} = 0.1$ (solid black lines in a,b,c). The dominant modes are: (a) $\lambda_y/\overline{\delta} \approx 1.5$, (b) $\lambda_y/\overline{\delta} \approx 1$, and (c) $\lambda_y/\overline{\delta} \approx 0.64$ ($\lambda_y/S \approx 2$).}
	\label{fig:pod_spiv_lydom}
	
	\includegraphics[height=4cm, width=13cm, keepaspectratio]{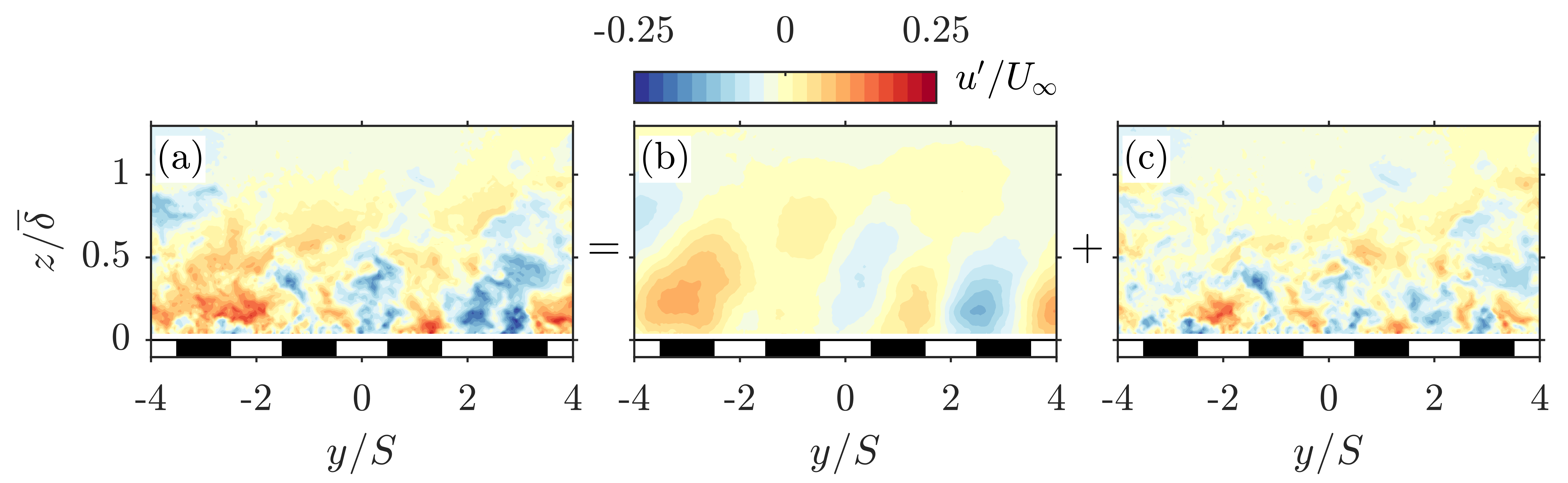}
	\caption{Turbulent velocity fluctuations $u'$ of case SR25 in $yz$-plane: (a) all resolved scales, (b) large scales (constructed from $u'$ POD modes with dominant $\lambda_y/\overline{\delta} > 0.64$), (c) small scales (constructed from $u'$ POD modes with dominant $\lambda_y/\overline{\delta} \leq 0.64$).}
	\label{fig:pod_spiv_recon}
\end{figure} 

Figure~\ref{fig:pod_spiv_lydom} shows POD modes of streamwise velocity fluctuation $\phi_{u'}^{i}$ for case SR25 in the $y$--$z$ plane. Three modes are selected: mode $i = 1$, 5, and 18, which contribute to 3.3\%, 2.2\%, and 0.9\% of the total kinetic energy, respectively. All three selected modes show a spanwise periodic pattern of low- and high-speed streaks (blue and red filled contours, respectively). Such streaks could be due to either LSM/VLSM or heterogeneity-induced secondary flows. To measure the `scale' of turbulence in each mode, the Fast Fourier Transform (FFT) is computed for each $\phi_{u'}$ at a reference wall-normal location $z/\overline{\delta} = 0.1$ (as close to the wall as the FOV permits, see solid black lines in figure~\ref{fig:pod_spiv_lydom}a--c). The 1-D power spectral density of $\phi_{u'}$ at $z/\overline{\delta} = 0.1$ is defined as
\begin{equation}
\Upphi_{\phi \phi} = \frac{2 c_n c_n^{\ast}}{\Updelta k_y}
\label{eq:psd_1d}
\end{equation}       
where $c_n$ is the $n$th Fourier coefficient, $c_n^{\ast}$ is its complex conjugate, and $k_y$ is the spanwise wavenumber. Figures~\ref{fig:pod_spiv_lydom}(d--f) show the 1-D energy spectra premultiplied by $k_y$ of mode 1, 5, and 18, respectively. These figures show that the spanwise periodic pattern observed in figure~\ref{fig:pod_spiv_lydom}(a--c) corresponds to a dominant spanwise wavelength $\lambda_y$, i.e. maximum $k_y \Upphi_{\phi \phi}$. Mode 1, for example, has the dominant $\lambda_y/\overline{\delta} \approx 1.5$, while the dominant wavelength for mode 5 and 18 are $\lambda_y/\overline{\delta} \approx 1$ and $\approx 0.64$, respectively. Dominant spanwise wavelengths are then computed for all 4800 resolved POD modes and separated into two groups: `large' scales (dominant $\lambda_y/\overline{\delta} > 0.64$, presumably related to LSM/VLSM) and `small' scales (dominant $\lambda_y/\overline{\delta} \leq 0.64 = 2S$ for case SR25 which will contain a mix of secondary flows due to spanwise heterogeneity as well as smaller scale non-stationary turbulent features). With this threshold for `large'/`small' scale separation, the POD modes associated with `large' and `small' scales make up 0.6\% and 99.4\% of all resolved modes, but contribute to 33.7\% and 66.3\% of the total kinetic energy, respectively. The fluctuating velocity fields are then reconstructed from both `large' and `small' scale POD modes in equation~\ref{eq:pod}. Figure~\ref{fig:pod_spiv_recon} shows an instantaneous fluctuating velocity field $u'$ of case SR25 ($S/\overline{\delta} = 0.32$) in the $yz$-plane, with figure~\ref{fig:pod_spiv_recon}(a) showing the broadband PIV measurement result (all resolved scales), figure~\ref{fig:pod_spiv_recon}(b) is the instantaneous $u'$ reconstructed only from POD modes whose dominant $\lambda_y/\overline{\delta} > 0.64$ (`large' scales), and figure~\ref{fig:pod_spiv_recon}(c) is reconstructed only from `small scales' ($\lambda_y/\overline{\delta} \leq 0.64$).

\begin{figure}
	\centering\includegraphics[height=7.75cm, width=13cm, keepaspectratio]{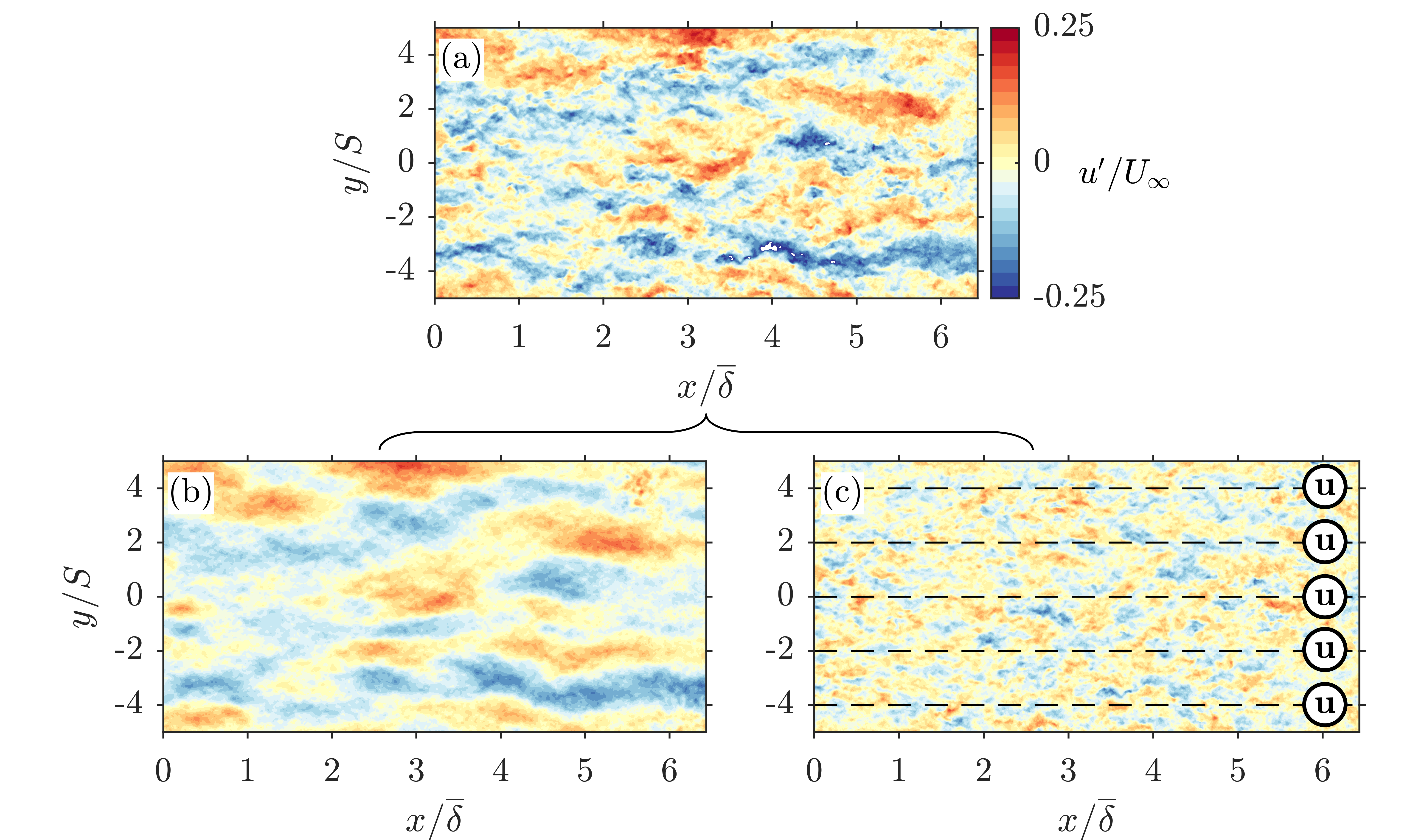}
	\caption{Turbulent velocity fluctuations $u'$ of case SR25 in $xy$-plane ($z/\overline{\delta} = 0.18$): (a) all resolved scales, (b) large scales (constructed from $u'$ POD modes with dominant $\lambda_y/\overline{\delta} > 0.64$), (c) small scales (constructed from $u'$ POD modes with dominant $\lambda_y/\overline{\delta} \leq 0.64$).}
	\label{fig:pod_wppiv_recon}
\end{figure}

\begin{figure}	
	\centering
	\includegraphics[height=4cm, width=12cm, keepaspectratio]{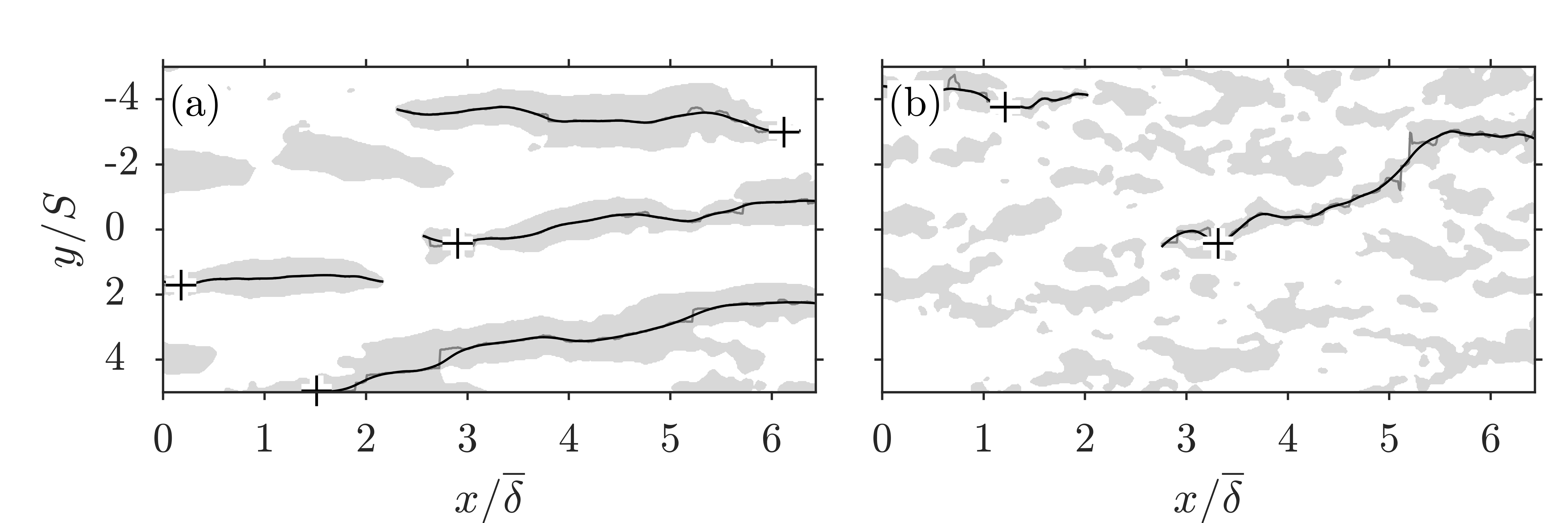}
	\caption{Detected low-speed structures $\widetilde{u}'$ for case SR25 ($z/\overline{\delta} = 0.18$): (a) large scales (constructed from $u'$ POD modes with dominant $\lambda_y/\overline{\delta} > 0.64$), (b) small scales (constructed from $u'$ POD modes with dominant $\lambda_y/\overline{\delta} \leq 0.64$). Grey coloured contours are the low-speed structures, $\widetilde{u}'/U_{\infty} < -0.03$. `+' marks the minima of a low-speed structure. The spines of the detected low-speed structures with length $\geq 2\overline{\delta}$ are shown in solid lines (from PIV data: \solidline[LightGray], smoothed: \solidline[black]). Dashed lines (\dashedline[black]) are the spanwise locations of the common flow up of the secondary flows (marked by \circled[u]).}
	\label{fig:pod_spines_z0.2}
	
	\centering
	\includegraphics[height=4.5cm, width=9cm, keepaspectratio]{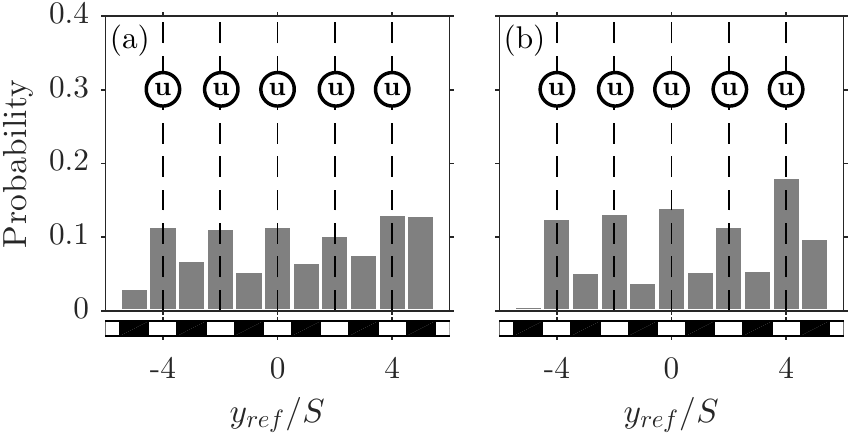}
	\caption{Histogram of the minima detected low-speed structures $y_{ref}$ for case SR25 ($z/\overline{\delta} = 0.18$): (a) large scales (figure~\ref{fig:pod_spines_z0.2}a) and (b) small scales (figure~\ref{fig:pod_spines_z0.2}b). Dashed lines (\dashedline[black]) are the spanwise locations of the common flow up of the secondary flows (marked by \circled[u]).}
	\label{fig:pod_hist_z0.2}
\end{figure} 

A similar strategy to separate turbulent scales related to LSM/VLSM and the secondary flows is also applied for the fluctuating velocity fields in the $x$--$y$ plane (WPPIV data) at $z/\overline{\delta} = 0.18$. This wall-normal location of the $xy$-plane is close to the centre of the secondary flows, which is estimated at $z/\overline{\delta} \approx 0.16$, $z/S \approx 0.5$. To obtain the `scales' of turbulent flows, FFT is once again computed for each $\phi_{u'}$ in both the $x$- and $y$-directions. The 2-D power spectral density of $\phi_{u'}$ in the $xy$-plane is defined as
\begin{equation}
\Upphi_{\phi \phi} = \frac{4 c_n c_n^{\ast}}{\Updelta k_x \Updelta k_y}
\label{eq:psd_2d}
\end{equation}
where $k_x$ is the streamwise wavenumber. The maxima of the premultiplied 2-D energy spectra of each POD mode $k_x k_y \Upphi_{\phi \phi}$ corresponds to the dominant streamwise and spanwise wavelength ($\lambda_x$ and $\lambda_y$) of each mode. It should be noted that the accuracy of this method to detect dominant $\lambda_x$ and $\lambda_y$ is limited by the streamwise and spanwise extent of the FOV. Similar to the POD modes of $u'$ in the $y$--$z$ plane, $\phi_{u'}$ in the $x$--$y$ plane are also separated into two groups based on the dominant $\lambda_y$: `large' scales (dominant $\lambda_y/\overline{\delta} > 0.64$) and `small' scales (dominant $\lambda_y/\overline{\delta} \leq 0.64 = 2S$). No such threshold for scale separation is applied for the dominant $\lambda_x$. With this threshold for `large'/`small' scale separation, the POD modes in the $x$--$y$ planes associated with `large' and `small' scales make up 6.3\% and 93.7\% of all resolved 1200 modes, but contribute to 32.8\% and 67.2\% of the total kinetic energy, respectively. Figure~\ref{fig:pod_wppiv_recon} illustrates the fluctuating velocity fields $u'$ reconstructed from both `large' and `small' scale POD modes in equation~\ref{eq:pod}. The conditional average of turbulent fluctuation $u'$ is computed at the minima and maxima of the low-speed structures in the $x$--$y$ plane. The same algorithm to detect `spines' of the structures and the minima/maxima in \S\ref{sub:sond_gg1} is also applied for SR25. However, for previously analysed cases where $S/\overline{\delta} \gg 1$ in \S\ref{sub:sond_gg1} the low-speed structures were separated into LSM/VLSM and secondary flows based on their spanwise locations. Here, where $S/\overline{\delta} \ll 1$, each instantaneous turbulent fluctuation field is separated into the `large' scales (presumably LSM/VLSM) and `small' scales (presumably containing the secondary flows) based on the dominant $\lambda_y$ of the POD modes. An example of detected low-speed structures for case SR25 ($z/\overline{\delta} = 0.18$) are shown in figure~\ref{fig:pod_spines_z0.2}(a) for the `large' scales and in figure~\ref{fig:pod_spines_z0.2}(b) for the `small' scales (gray filled contours are $\widetilde{u}'/U_{\infty} < -0.03$). Due to the decreasing length of the structures detected as the `small' scales (figure~\ref{fig:pod_wppiv_recon}c and figure~\ref{fig:pod_spines_z0.2}b), only structures whose length is $\geq 2\overline{\delta}$ are included in the computation (this criterion for low-speed structure detection is lower than that in \S\ref{sub:sond_gg1}).  Figure~\ref{fig:pod_hist_z0.2} shows the histogram of the $y$-minima of the detected low-speed structures constructed from large (figure~\ref{fig:pod_hist_z0.2}a) and small scales (figure~\ref{fig:pod_hist_z0.2}b). The smaller scales exhibit a phase-locking behaviour of the structures, with higher possibility of the spines detected over the common flow up. This phase-locking behaviour is still apparent (to lesser extent) in the larger scales since $\widetilde{u}'$ is used for detection (cf. figure~\ref{fig:spines}f for spine detection without scale separation). With this criterion for low-speed structure detection, 40\% of the detected streaks belong to the small scales, and 60\% belong to the large scales.


\begin{figure}
	\centering
	\includegraphics[height=12.1cm, width=13cm, keepaspectratio]{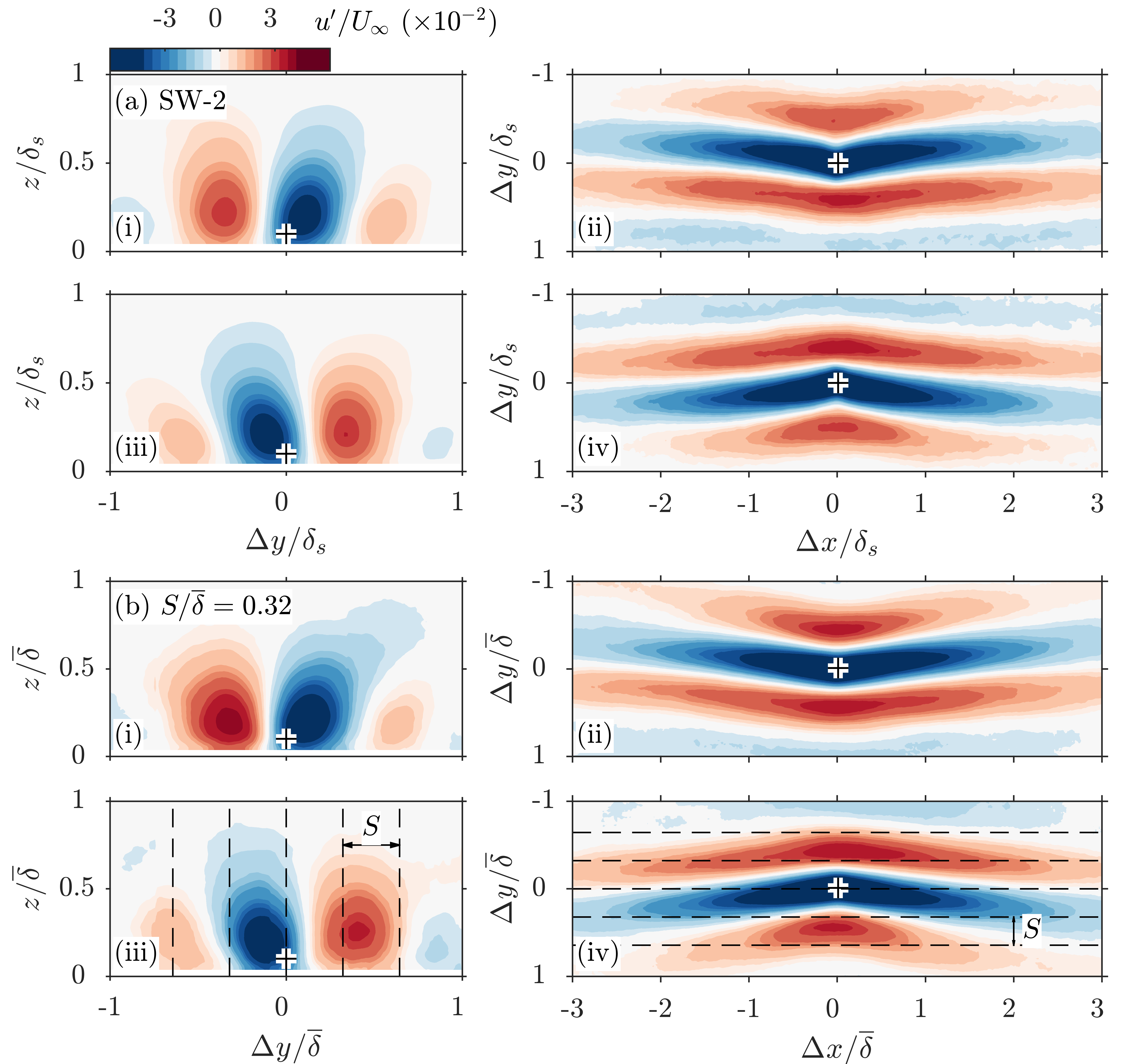}
	\caption{Contours of conditionally averaged large-scale turbulent fluctuation $u'$ (constructed from $u'$ POD modes with dominant $\lambda_y/\overline{\delta} > 0.64$ or $\lambda_y/\delta_s > 0.64$) for (a) reference case SW-2 at $z/\delta_s = 0.24$ and (b) SR25 at $z/\overline{\delta} = 0.18$. Conditions: (i) $\widetilde{u}' < 0$ and $v' > 0$, (ii) $\widetilde{u}' < 0$ and $v' < 0$ at `+', (iii) minima and (iv) maxima of detected low-speed structures (at `+', $\widetilde{u}'/U_{\infty} < -0.03$) illustrated in figure~\ref{fig:pod_spines_z0.2}(a) (refer to \S\ref{sub:sond_eq1} for the method used to extract the `spines' of low-speed structures). In (b)(iii) and (iv), dashed lines (\dashedline[black]) show the width of the roughness strips $S$.}
	\label{fig:pod_condave_z0.2_lp}
\end{figure} 


Conditional averages of $u'$ are computed in the $y$--$z$ and $x$--$y$ planes for both `large' and `small' scales. In the $y$--$z$ plane, the condition point of $u'$ (`+' in figure~\ref{fig:pod_condave_z0.2_lp}i and iii) is at $z/\overline{\delta} = 0.1$ (as close to the wall as the FOV permits) and at the spanwise locations of common flow up (marked by \circled[u], centre of the smooth strips for this case) for consistency with the previous analysis in \S\ref{sub:sond_eq1} and \S\ref{sub:sond_gg1} (for `larger' scales, the resulting conditionally averaged structures are insensitive to the spanwise location of the condition vector). The conditions for the $y$--$z$ plane are as follows: $\widetilde{u}' < 0$ (low-speed streaks) and $v' > 0$ or $\widetilde{u}' < 0$ (low-speed streaks) and $v' < 0$. In the $x$--$y$ plane, the condition point is at the $y$-minima and $y$-maxima of the spines fitted to the detected low-speed streaks, which corresponds to $v' > 0$ and $v' < 0$ (see figure~\ref{fig:condave_uv} for an example). To provide a valid comparison, the same POD and conditions are also applied to the reference smooth wall case SW-2 with appropriate scaling ($\delta_s$ instead of $\overline{\delta}$). For the smooth case, the velocity field is also separated into `large' (POD modes with dominant $\lambda_y/\delta_s > 0.64$) and `small' scales ($\lambda_y/\delta_s \leq 0.64$). In the smooth $y$--$z$ plane, the condition point of $u'$ is at $z/\delta_s = 0.1$ and spanwise homogeneity is assumed. For the smooth $x$--$y$ plane, only structures with length $\geq 2\delta_s$ are included and spanwise homogeneity is also assumed. With this criterion for low-speed structure detection, 13\% of the detected streaks belong to the small scales, and 87\% belong to the large scales.

Figure~\ref{fig:pod_condave_z0.2_lp}(a,b) show the contours of conditionally averaged `large'-scale $u'$ for the reference smooth wall case SW-2 and case SR25 ($S/\overline{\delta} = 0.32$), respectively. Conditionally averaged $u'$ in the $y$--$z$ plane is shown in subplot (i) and (iii).  The $x$--$y$ plane (subplot ii and iv) is located at $z/\delta_s = 0.24$ for SW-2 and $z/\overline{\delta} = 0.18$ for SR25 (close to the centre of the secondary flows). The contours show the structures leaning to the left and right, similar to those shown in \S\ref{sub:sond_eq1} and \S\ref{sub:sond_gg1}. It is interesting to note, however, that when only `large' scales (i.e. scales that are larger than the spanwise heterogeneity) are considered, the structures of SW-2 and SR25 are identical, further suggesting that in the case of spanwise heterogeneity where the roughness $S/\overline{\delta} \ll 1$, scales that are $\gtrsim S$ can be considered as the naturally occurring turbulent structures and they exist in a relatively unchanged form compared to those observed in the reference smooth wall. This is also in line with Townsend's outer layer similarity for rough walls. As $S/\overline{\delta}$ becomes very small, the wall condition approaches large scale homogeneity, and we might expect that beyond the roughness sublayer ($z> S$, see \citealp{chan2018}), outer layer similarity should be preserved. Locking of large-scales now would be minimal, and it would only be scales that were close to $\lambda_y = 2S$ that we would perhaps expect to be locked by the spanwise heterogeneity.

\begin{figure}
	\centering
	\includegraphics[height=17.1cm, width=13cm, keepaspectratio]{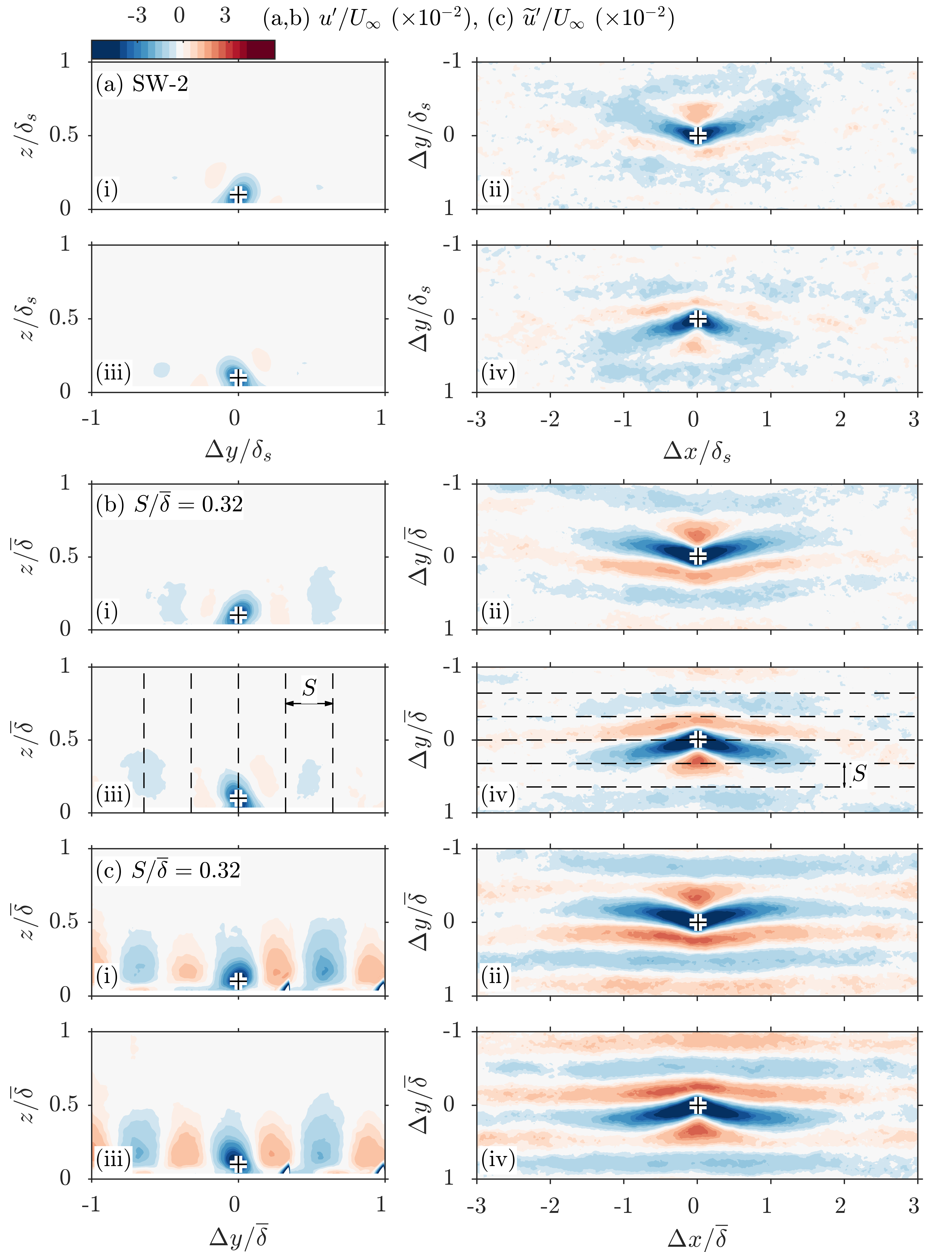}
	\caption{Contours of conditionally averaged small-scale turbulent fluctuation $u'$ (constructed from $u'$ POD modes with dominant $\lambda_y/\overline{\delta} \leq 0.64$ or $\lambda_y/\delta_s \leq 0.64$) for (a) reference case SW-2 at $z/\delta_s = 0.24$, (b) SR25 at $z/\overline{\delta} = 0.18$, and (c) conditionally averaged $\widetilde{u}'$ (small-scale $u'$ supplemented with $\widetilde{U}$) for case SR25 at $z/\overline{\delta} = 0.18$. Conditions: (i) $\widetilde{u}' < 0$ and $v' > 0$, (ii) $\widetilde{u}' < 0$ and $v' < 0$ at `+', (iii) minima and (iv) maxima of detected low-speed structures (at `+', $\widetilde{u}'/U_{\infty} < -0.03$) illustrated in figure~\ref{fig:pod_spines_z0.2}(b) (refer to \S\ref{sub:sond_eq1} for the method used to extract the `spines' of low-speed structures). In (b)(iii) and (iv), dashed lines (\dashedline[black]) show the width of the roughness strips $S$.}
	\label{fig:pod_condave_z0.2_hp}
\end{figure}

Figure~\ref{fig:pod_condave_z0.2_hp}(a,b) show the contours of conditionally averaged `small'-scale $u'$ for the reference smooth wall case SW-2 and case SR25 ($S/\overline{\delta} = 0.32$), respectively. In the $y$--$z$ plane, the structures for both SW-2 and SR25 are capped in the wall-normal direction at $z/\overline{\delta} \lesssim 0.32$ ($z/\delta_s \lesssim 0.32$), as suggested by \cite{chan2018}, who observed that the roughness sublayer height is approximately half of the spanwise wavelength of the roughness. However, a clear spanwise periodic pattern is observed in the the SR25 case (figure~\ref{fig:pod_condave_z0.2_hp}b, subplot i and iii), which is less apparent in the smooth wall reference (figure~\ref{fig:pod_condave_z0.2_hp}a, subplot i and iii). The wavelength of this spanwise periodic pattern is $2S$ (which is also the spanwise roughness wavelength $\Lambda$), as shown by the dashed lines in figure~\ref{fig:pod_condave_z0.2_hp}(b)(iii). A slight misalignment is attributed to the condition of $v'$, such that structures lean slightly to the left and right. In the $xy$-plane, a similar pattern is also observed. The pronounced spanwise periodicity for the SR25 case is largely absent for the smooth wall reference. Furthermore, the streaks for case SR25 appear to be elongated in $x$ (approximately 3--$4\overline{\delta}$ long, figure~\ref{fig:pod_condave_z0.2_hp}(b)ii and iv) to a greater extent than those observed for SW-2 ($\sim 2\delta_s$). This length is similar to the streamwise wavelength associated with the additional energy spectra peak \citep[see][figure 14f]{wangsawijaya2020}, which has been previously associated with meandering of the secondary flows. It is important to note that the conditional averages shown in figure~\ref{fig:pod_condave_z0.2_hp} are for $u'$, the convecting non-stationary turbulent component. A locked fluctuation such as this is distinct from the stationary secondary flows (which would appear in $\widetilde{u}'$, as shown in figure~\ref{fig:pod_condave_z0.2_hp}c). A possible explanation for these elongated streaks would be a meandering behaviour of the secondary flows occurring at the interface between the rough and smooth strips. The pronounced spanwise periodicity exhibited for case SR25 in figure~\ref{fig:pod_condave_z0.2_hp} is more difficult to explain, but implies cross-talk between adjacent secondary flows such their meandering is in-phase. The results presented in figures~\ref{fig:pod_condave_z0.2_lp} and~\ref{fig:pod_condave_z0.2_hp} strongly suggest that the secondary flows and naturally occurring $\overline{\delta}$-scaled LSM/VLSM co-exist in the case where $S/\overline{\delta} \ll 1$. With $S$-scaled features residing close to the wall ($z < S$) that are locked to the phase of the spanwise heterogeneity, and with superimposed large-scale features that are virtually unchanged from those occurring over smooth surfaces, exhibiting no spanwise locking or alteration due to the heterogeneity.

\section{Conclusions}

We conduct an analysis of secondary flows and perturbed turbulent boundary layers resulting from spanwise varying surface conditions within a range of spanwise half-wavelength $0.32 \leq S/\overline{\delta} \leq 3.63$. In \cite{wangsawijaya2020}, it was suggested that $S/\overline{\delta}$ governs the size and strength of the secondary flows. The latter is maximum for intermediate cases ($S/\overline{\delta} \approx 1$), which suggests two possible scenarios: either (i) that these cases induce the strongest secondary streamwise vortices or (ii) if secondary flows may be considered as spatially locked turbulent structures, that these are more effectively locked in place for these heterogeneous wavelengths. Present results suggest that (ii) is viable. Following this scenario, we might consider the secondary flows (instantaneously) as spanwise-locked large-scale turbulent structures. The 1-D spanwise energy spectra (figure~\ref{fig:spiv_spectra} and~\ref{fig:spiv_spectra_int}) suggests that there is an interplay between the stationary secondary flows and the turbulent fluctuations, where energy of the first might leech into the latter, and this is also governed by $S/\overline{\delta}$. In the intermediate cases ($S/\overline{\delta} \approx 1$) where the secondary flows are space filling (hence, effectively locked), the energy deposited into the stationary component and turbulent fluctuations reach maximum and minimum, respectively. This might further suggest that $S/\overline{\delta}$ determines the efficacy of roughness wavelengths in locking the secondary flows/turbulent structures in place.
 
In addition to being spanwise locked, it was also suggested in \cite{wangsawijaya2020} that the secondary flows exhibit a quantifiable streamwise unsteadiness, which was interpreted as a statistical evidence for meandering with a preferred streamwise wavelength when $S/\overline{\delta} \approx 1$ \citep[see figure 15, 18, and 19]{wangsawijaya2020}. In the current study we reveal the strongly meandering structures through conditional averaging of $u'$ for both SR cases and the reference smooth-wall case. The resulting structures for both of these cases are strongly reminiscent of the streak-vortex instability model proposed in \cite{jeong1997,waleffe2001,flores2010,cossu2017} and \cite{degiovanetti2017} (figure~\ref{fig:condave_3d}).        

In the limits where $S/\overline{\delta} \gg 1$ and $S/\overline{\delta} \ll 1$, it is possible to examine the coexistence of secondary flows and the naturally occuring large-scale structures. In the cases where $S/\overline{\delta} \gg 1$ both the secondary flows and large-scale structures are $\delta$-scaled. The first are locked about the roughness transition and still carry the imprint of meandering secondary flows, although not as strong as the intermediate cases, while the latter occur farther from the transition and are identical to those observed in the reference smooth wall case (figure~\ref{fig:condave_sr250}). In the other limit where $S/\overline{\delta} \ll 1$, secondary flows and large-scale structures scale on $S$ and $\delta$, respectively. Using a POD-based filter, each velocity field snapshot can be separated into scales $ > S$ or $\leq S$. The conditional average of $u'$ of the larger scales show that the structures of turbulence in the limiting case ($S/\overline{\delta} \ll 1$) and the reference smooth wall are identical (figure~\ref{fig:pod_condave_z0.2_lp}), suggesting that structures with spanwise scales $> 2S$ are unaffected by the heterogeneity. The smaller scales for the spanwise heterogeneous cases, however, exhibit the imprint of secondary flows leeching into $u'$, characterised as long meandering spanwise alternating streaks that are locked and aligned with the roughness strips. One outstanding question that remains unanswered in the present study is the cause of the prominent meandering of the turbulent structures, which is only observed when $S/\overline{\delta} \approx 1$. This will be an interesting question to address in the future.

\backsection[Supplementary data]{Not available.}


\backsection[Funding]{This research is supported by the Australian Research Council Discovery Project (DP160102279) and the Office of Naval Research (BRC N00014-17-1-2307).}

\backsection[Declaration of interests]{The authors report no conflict of interest.}

\backsection[Data availability statement]{The data that support the findings of this study are available from the corresponding author (D.D.W.).}

\backsection[Author ORCID]{D. D. Wangsawijaya, https://orcid.org/
0000-0002-7072-4245; N. Hutchins, https://orcid.org/
0000-0003-1599-002X}

\backsection[Author contributions]{D.D.W. designed, performed all experiments, and analysed the data with the supervision of N.H. D.D.W and N.H. wrote the manuscript.}

\appendix
\section{Stationary vs. non-stationary component of the secondary flows}
\label{app}

\begin{figure}
	\centering
	\includegraphics[height=4.5cm, width=13cm, keepaspectratio]{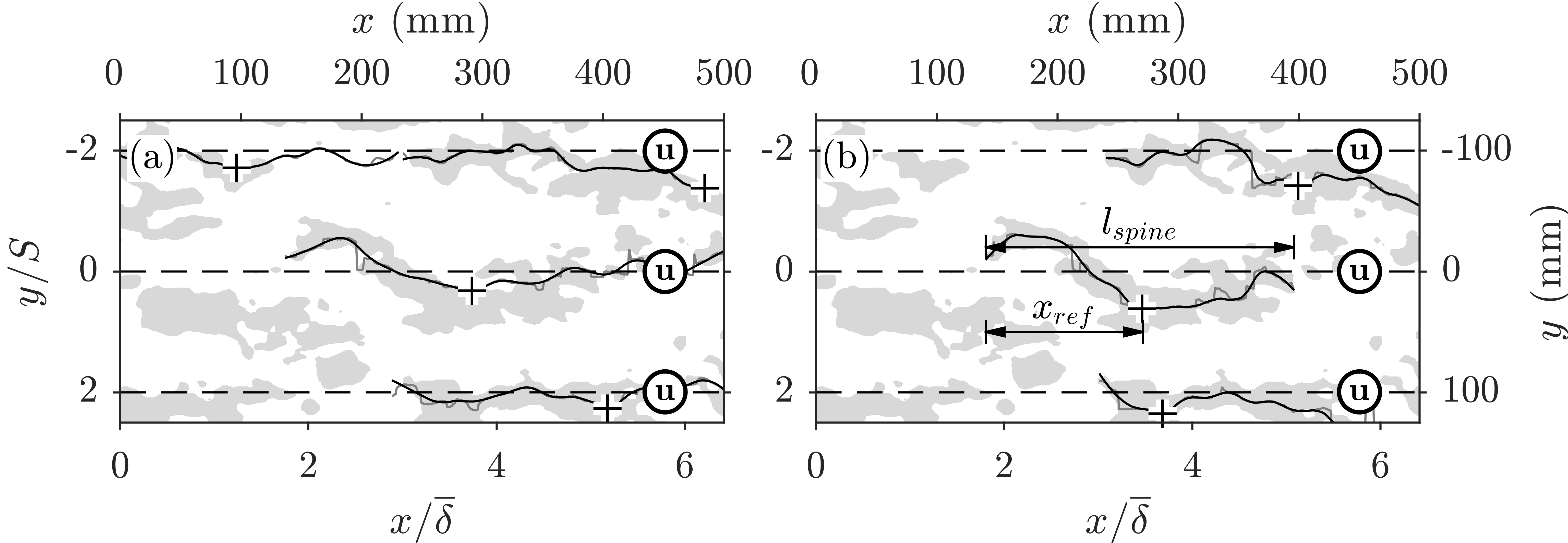}
	\caption{Detected low-speed structures for case SR50 ($S/\overline{\delta} = 0.62$) from a WPPIV snapshot. Gray-coloured contours are: (a) $\widetilde{u}'/U_{\infty} < -0.03$ and (b) $u'/U_{\infty} < -0.03$. `+' marks the minima of a low-speed structure. The spines of the detected low-speed structures are shown in solid lines (from PIV data: \solidline[LightGray], smoothed: \solidline[black]). Dashed lines (\dashedline[black]) are the spanwise locations of the common flow up of the secondary flows (marked by \circled[u], see figure~\ref{fig:U_decomposed}(g) for these locations in the $yz$-plane).}
	\label{fig:spines_utildevsuprime}
\end{figure}

\begin{figure}
	\centering
	\includegraphics[height=6.5cm, width=13cm, keepaspectratio]{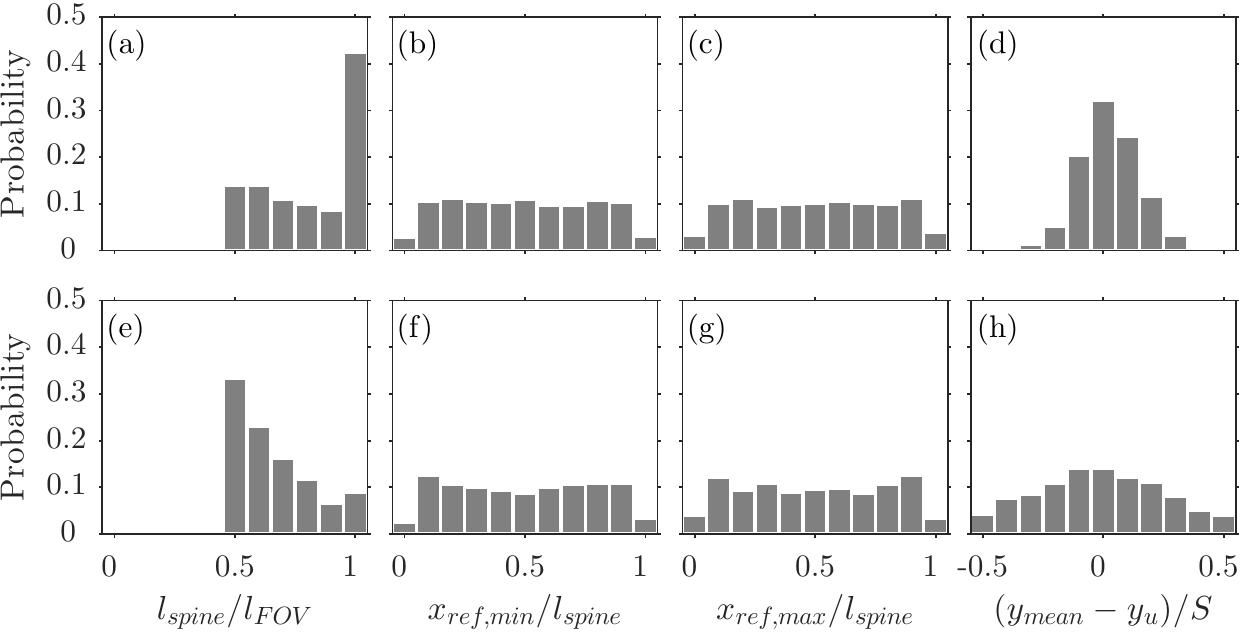}
	\caption{Histograms of the low-speed structures for case SR50 ($S/\overline{\delta} = 0.62$), detected in the fluctuating velocity field: (a--d) $\widetilde{u}'$ and (e--h) $u'$. (a,e) length of spines of the detected structures relative to the FOV, (b,f) streamwise position of the detected minima and (c,g) maxima relative to the length of the spines, (d,h) spanwise position of the spines relative to the common flow up.}
	\label{fig:hist_utildevsuprime}
\end{figure}

Figure~\ref{fig:spines_utildevsuprime} compares the extracted spines from an instantaneous velocity field $\widetilde{u}' \equiv \widetilde{U} + u'$ (figure~\ref{fig:spines_utildevsuprime}a) and $u'$ (figure~\ref{fig:spines_utildevsuprime}b) of case SR50 ($S/\overline{\delta} = 0.62$). The spine extraction method for both $\widetilde{u}'$ and $u'$ velocity fields is the same: both are filtered with a box filter of $0.1\overline{\delta} \times 0.1\overline{\delta}$ size, the gray-coloured contours in figure~\ref{fig:spines_utildevsuprime} are $\widetilde{u}'/U_{\infty} < -0.03$ and $u'/U_{\infty} < -0.03$, respectively, and only structures longer than or equal to $3\overline{\delta}$ are kept for the analysis. For comparison between $\widetilde{u}'$ and $u'$, we introduce three parameters related to the extracted spines: the length of the spine $l_{spine}$ relative to the streamwise extent of the FOV $l_{FOV}$, the location of the minima/maxima $x_{ref}$ relative to the spine, and the location of the spine relative to the common flow up (\circled[u] in figure~\ref{fig:spines_utildevsuprime}). Figure~\ref{fig:hist_utildevsuprime}(a,e) shows the histogram of the length of the extracted spines $l_{spine}$ from $\widetilde{u}'$ and $u'$ velocity fields, respectively. When the $\widetilde{u}'$ field is used for extraction, the spines are likely to extend across the streamwise extent of the FOV ($l_{spine}/l_{FOV} = 1$). The spines are mostly shorter ($l_{spine}/l_{FOV} = 0.5$) when $u'$ is used instead of $\widetilde{u}'$. This is expected since the stationary component of the secondary flows $\widetilde{U}$ has an infinite mode in $x$, which is present for $\widetilde{u}'$ but removed for $u'$. The histograms of the streamwise location $x_{ref}$ of the minima of the extracted structures relative to the start of the detected structure (see `+' and the definition of $x_{ref}$ in figure~\ref{fig:spines_utildevsuprime}) from $\widetilde{u}'$ and $u'$ fields are shown in figure~\ref{fig:hist_utildevsuprime}(b,f), respectively, while the histograms for the maxima are shown in figure~\ref{fig:hist_utildevsuprime}(c,g). The histograms illustrate the location (of the minima and maxima) relative to the extracted spines $x_{ref}/l_{spine}$. Here, $x_{ref}/l_{spine} = 0$ means that the minima/maxima is located at upstream end of the spine, while $x_{ref}/l_{spine} = 1$ means that it is located at the downstream end. Figure~\ref{fig:hist_utildevsuprime}(b,c,f,g) shows that $x_{ref}$ is roughly evenly distributed across 10--90\% length of the spines when either $\widetilde{u}'$ or $u'$ is used for extraction. Lastly, we also examine the spanwise locking of the low-speed structures. We compute the distance between the $y$ location of the spines $y_{mean}$ and the location of common flow up $y_{up}$ (dashed black lines in figure~\ref{fig:spines_utildevsuprime}). Here, $y_{mean}$ is the $y$-average of all extracted spines (solid black lines in figure~\ref{fig:spines_utildevsuprime}). Figure~\ref{fig:hist_utildevsuprime}(d,h) shows the histograms of the calculated distance between the spines and common flow up normalised by $S$ when $\widetilde{u}'$ and $u'$ field are used for extraction, respectively. Here, $(y_{mean}-y_u)/S = 0$ is the center of a smooth strip (which coincides with the common flow up) and $(y_{mean}-y_u)/S \pm 0.5$ is the interface between a smooth strip and adjacent rough strips (figure~\ref{fig:U_decomposed}). The histograms show that the extracted spines from $\widetilde{u}'$ are more likely to be locked on the common flow up ($y_u$) compared to those extracted from $u'$, which is expected since $\widetilde{U}$ is a minima at $y_u$, and hence we are much more likely to detect a minima in $\widetilde{u}' \equiv \widetilde{U} + u'$ at this location.


\begin{figure}
	\centering
	\includegraphics[height=5.9cm, width=12.6cm, keepaspectratio]{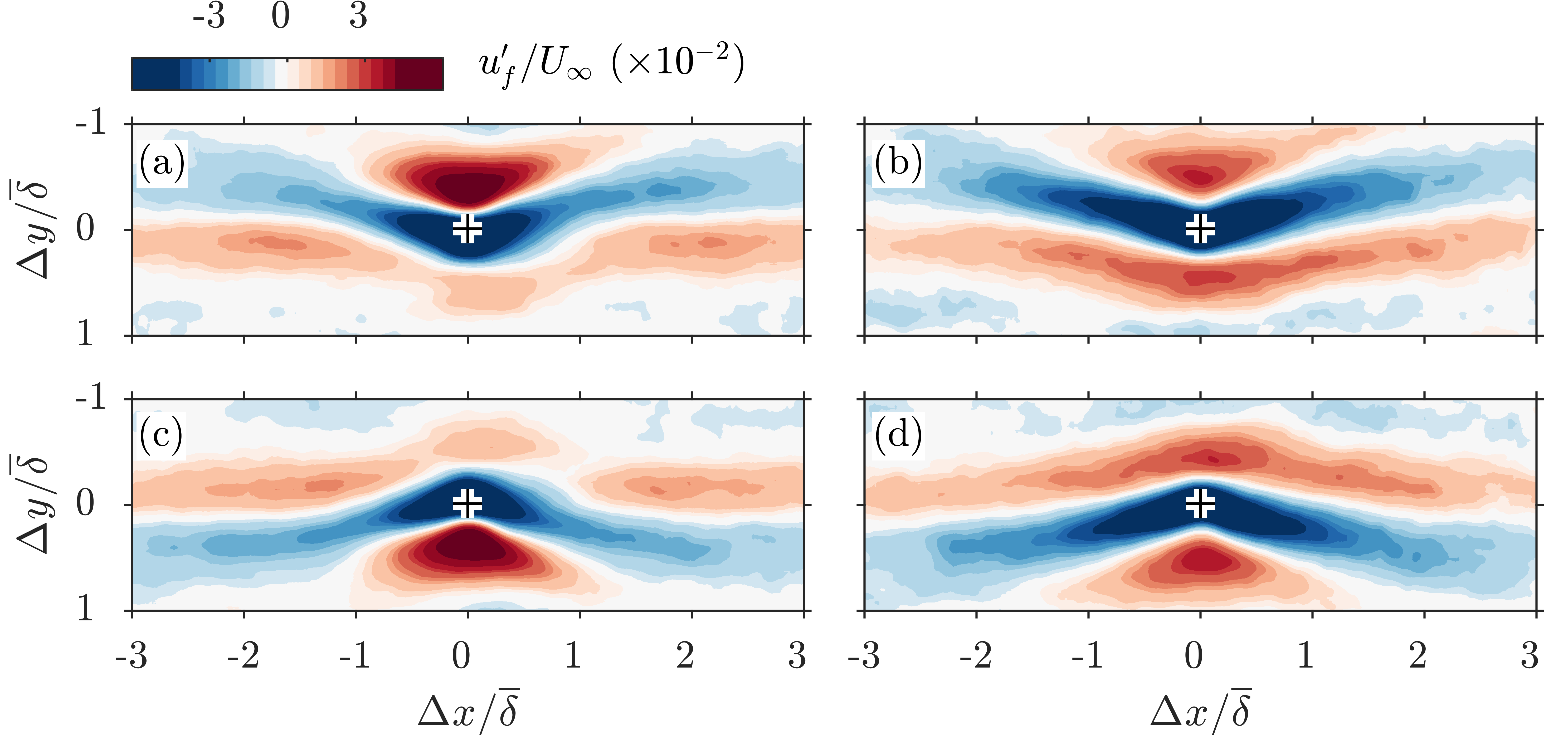}
	\caption{Contours of filtered turbulent fluctuation $u_f'$ conditionally averaged at (a,b) the minima and (c,d) maxima of the detected low-speed structures for case SR50 ($S/\overline{\delta} = 0.62$). Structures are detected in the fluctuating velocity field: (a,c) $\widetilde{u}'$ and (b,d) $u'$, as illustrated in figure~\ref{fig:spines_utildevsuprime}.}
	\label{fig:condave_utildevsuprime}
\end{figure}

We repeat the computation of the conditionally averaged fluctuating velocity field previously introduced in \S\ref{sub:sond_eq1} at the minima and maxima of the spines extracted from $u'$ field (instead of $\widetilde{u'}$). It should be noted that the conditional average is computed for filtered turbulent fluctuation $u_f'$, but the condition points differ between the spines extracted from $\widetilde{u}'$ and $u'$ field (see `+' in figure~\ref{fig:spines_utildevsuprime}). Contours of conditionally averaged $u_f'$ of case SR50 ($S/\overline{\delta} = 0.62$) are shown in figure~\ref{fig:condave_utildevsuprime}(a,c) for condition points extracted from $\widetilde{u}'$ and figure~\ref{fig:condave_utildevsuprime}(b,d) from $u'$. Both cases show blue-filled contours (low-speed structures) flanked by red-filled contours (high-speed structures), but the strength of flanking high speed events seems stronger and more symmetrically arranged about the low speed streak for $u'$ (figure~\ref{fig:condave_utildevsuprime}b,d) than $\widetilde{u}'$ (figure~\ref{fig:condave_utildevsuprime}a,c). Further, the amplitude of meandering for $u'$ is only slightly lower than that of $\widetilde{u}'$. This suggests that the choice of $\widetilde{u}'$ or $u'$ makes very little difference in the resulting conditionally averaged structures, and explains the very marginal influence of choice of condition vector evident in figure~\ref{fig:yhat}(b). 

\bibliographystyle{jfm}
\bibliography{{817596_bibliography}}

\begin{thebibliography}{49}
\expandafter\ifx\csname natexlab\endcsname\relax\def\natexlab#1{#1}\fi
\def\au#1{#1} \def\ed#1{#1} \def\yr#1{#1}\def\at#1{#1}\def\jt#1{\textit{#1}}
  \def\bt#1{#1}\def\bvol#1{\textbf{#1}} \def\vol#1{#1} \def\pg#1{#1}
  \def\publ#1{#1}\def\arxiv#1{#1}\def\org#1{#1}\def\st#1{\textit{#1}}

\bibitem[Adrian {\em et~al.\/}(2000)Adrian, Meinhart \& Tomkins]{adrian2000}
{\sc \au{Adrian, R.~J.}, \au{Meinhart, C.~D.} \& \au{Tomkins, C.~D.}} \yr{2000}
   \at{Vortex organization in the outer region of the turbulent boundary
  layer}.  \jt{J. Fluid Mech.}  \bvol{422},  \pg{1--54}.

\bibitem[Awasthi \& Anderson(2018)]{awasthi2018}
{\sc \au{Awasthi, A.} \& \au{Anderson, W.}} \yr{2018}  \at{Numerical study of
  turbulent channel flow perturbed by spanwise topographic heterogeneity:
  Amplitude and frequency modulation within low- and high-momentum pathways}.
  \jt{Phys. Rev. Fluids}  \bvol{3},  \pg{044602}.

\bibitem[Balakumar \& Adrian(2007)]{balakumar2007}
{\sc \au{Balakumar, B.~J.} \& \au{Adrian, R.~J.}} \yr{2007}  \at{Large and
  very-large-scale motions in channel and boundary-layer flows}.  \jt{Phil.
  Trans. R. Soc. A}  \bvol{365},  \pg{665--681}.

\bibitem[Chan {\em et~al.\/}(2018)Chan, Chung, Hutchins \& Ooi]{chan2018}
{\sc \au{Chan, L.}, \au{Chung, M. MacDonald~D.}, \au{Hutchins, N.} \& \au{Ooi,
  A.}} \yr{2018}  \at{Secondary motion in turbulent pipe flow with
  three-dimensional roughness}.  \jt{J. Fluid Mech.}  \bvol{854},  \pg{5--33}.

\bibitem[Chung {\em et~al.\/}(2018)Chung, Monty \& Hutchins]{chung2018}
{\sc \au{Chung, D.}, \au{Monty, J.~P.} \& \au{Hutchins, N.}} \yr{2018}
  \at{Similarity and structure of wall-turbulence with lateral wall shear
  stress variations}.  \jt{J. Fluid Mech.}  \bvol{847},  \pg{591--613}.

\bibitem[Coceal {\em et~al.\/}(2006)Coceal, Thomas, Castro \&
  Belcher]{coceal2006}
{\sc \au{Coceal, O.}, \au{Thomas, T.~G.}, \au{Castro, I.~P.} \& \au{Belcher,
  S.~E.}} \yr{2006}  \at{Mean flow and turbulence statistics over groups of
  urban-like cubical obstacles}.  \jt{Bound.-Layer Meteorol.}  \bvol{121},
  \pg{491--519}.

\bibitem[Cossu \& Hwang(2017)]{cossu2017}
{\sc \au{Cossu, C.} \& \au{Hwang, Y.}} \yr{2017}  \at{Self-sustaining processes
  at all scales in wall-bounded turbulent shear flows}.  \jt{Phil. Trans. R.
  Soc. A}  \bvol{375},  \pg{20160088}.

\bibitem[Dennis \& Nickels(2011{\natexlab{{\em a\/}}})]{dennis2011a}
{\sc \au{Dennis, D.~J.~C.} \& \au{Nickels, T.~B.}} \yr{2011{\natexlab{{\em
  a\/}}}}  \at{Experimental measurement of large-scale three-dimensional
  structures in a turbulent boundary layer. \uppercase{P}art 1.
  \uppercase{V}ortex packets}.  \jt{J. Fluid Mech.}  \bvol{673},
  \pg{180--217}.

\bibitem[Dennis \& Nickels(2011{\natexlab{{\em b\/}}})]{dennis2011b}
{\sc \au{Dennis, D.~J.~C.} \& \au{Nickels, T.~B.}} \yr{2011{\natexlab{{\em
  b\/}}}}  \at{Experimental measurement of large-scale three-dimensional
  structures in a turbulent boundary layer. \uppercase{P}art 2.
  \uppercase{L}ong structures}.  \jt{J. Fluid Mech.}  \bvol{673},
  \pg{218--244}.

\bibitem[Elsinga {\em et~al.\/}(2010)Elsinga, Adrian, Oudheusden \&
  Scarano]{elsinga2010}
{\sc \au{Elsinga, G.~E.}, \au{Adrian, R.~J.}, \au{Oudheusden, B.~W.~Van} \&
  \au{Scarano, F.}} \yr{2010}  \at{Three-dimensional vortex organization in a
  high-reynolds-number supersonic turbulent boundary layer}.  \jt{J. Fluid
  Mech.}  \bvol{644},  \pg{35--60}.

\bibitem[Finnigan(2000)]{finnigan2000}
{\sc \au{Finnigan, J.}} \yr{2000}  \at{Turbulence in plant canopies}.
  \jt{Annu. Rev. Fluid Mech.}  \bvol{32},  \pg{519--571}.

\bibitem[Flores \& Jim\'{e}nez(2010)]{flores2010}
{\sc \au{Flores, O.} \& \au{Jim\'{e}nez, J.}} \yr{2010}  \at{Hierarchy of
  minimal flow units in the logarithmic layer}.  \jt{Phys. Fluids}
  \bvol{22}~(7),  \pg{071704}.

\bibitem[Ganapathisubramani {\em et~al.\/}(2005)Ganapathisubramani, Hutchins,
  Hambleton, Longmire \& Marusic]{ganapathisubramani2005}
{\sc \au{Ganapathisubramani, B.}, \au{Hutchins, N.}, \au{Hambleton, W.~T.},
  \au{Longmire, E.~K.} \& \au{Marusic, I.}} \yr{2005}  \at{Investigation of
  large-scale coherence in a turbulent boundary layer using two-point
  correlations}.  \jt{J. Fluid Mech.}  \bvol{524},  \pg{57--80}.

\bibitem[Ganapathisubramani {\em et~al.\/}(2003)Ganapathisubramani, Longmire \&
  Marusic]{ganapathisubramani2003}
{\sc \au{Ganapathisubramani, B.}, \au{Longmire, E.~K.} \& \au{Marusic, I.}}
  \yr{2003}  \at{Characteristics of vortex packets in turbulent boundary
  layers}.  \jt{J. Fluid Mech.}  \bvol{478},  \pg{35--46}.

\bibitem[de~Giovanetti {\em et~al.\/}(2017)de~Giovanetti, Sung \&
  Hwang]{degiovanetti2017}
{\sc \au{de~Giovanetti, M.}, \au{Sung, H.~J.} \& \au{Hwang, Y.}} \yr{2017}
  \at{Streak instability in turbulent channel flow: the seeding mechanism of
  large-scale motions}.  \jt{J. Fluid Mech.}  \bvol{832},  \pg{483--513}.

\bibitem[Guala {\em et~al.\/}(2006)Guala, Hommema \& Adrian]{guala2006}
{\sc \au{Guala, M.}, \au{Hommema, S.~E.} \& \au{Adrian, R.~J.}} \yr{2006}
  \at{Large-scale and very-large-scale motions in turbulent pipe flow}.  \jt{J.
  Fluid Mech.}  \bvol{564},  \pg{267--285}.

\bibitem[Hutchins {\em et~al.\/}(2012)Hutchins, Chauhan, Marusic, Monty \&
  Klewicki]{hutchins2012}
{\sc \au{Hutchins, N.}, \au{Chauhan, K.}, \au{Marusic, I.}, \au{Monty, Jason}
  \& \au{Klewicki, J.}} \yr{2012}  \at{Towards reconciling the large-scale
  structure of turbulent boundary layers in the atmosphere and laboratory}.
  \jt{Bound.-Layer Meteorol.}  \bvol{145},  \pg{273--306}.

\bibitem[Hutchins {\em et~al.\/}(2005)Hutchins, Hambleton \&
  Marusic]{hutchins2005}
{\sc \au{Hutchins, N.}, \au{Hambleton, W.~T.} \& \au{Marusic, I.}} \yr{2005}
  \at{Inclined cross-stream stereo particle image velocimetry measurements in
  turbulent boundary layers}.  \jt{J. Fluid Mech.}  \bvol{541},  \pg{21--54}.

\bibitem[Hutchins \& Marusic(2007{\natexlab{{\em a\/}}})]{hutchins2007a}
{\sc \au{Hutchins, N.} \& \au{Marusic, I.}} \yr{2007{\natexlab{{\em a\/}}}}
  \at{Evidence of very long meandering features in the logarithmic region of
  turbulent boundary layers}.  \jt{J. Fluid Mech.}  \bvol{579},  \pg{1--28}.

\bibitem[Hutchins \& Marusic(2007{\natexlab{{\em b\/}}})]{hutchins2007b}
{\sc \au{Hutchins, N.} \& \au{Marusic, I.}} \yr{2007{\natexlab{{\em b\/}}}}
  \at{Large-scale influences in near-wall turbulence}.  \jt{Phil. Trans. R.
  Soc. A}  \bvol{365},  \pg{647--664}.

\bibitem[Jeong {\em et~al.\/}(1997)Jeong, Hussain, Schoppa \& Kim]{jeong1997}
{\sc \au{Jeong, J.}, \au{Hussain, F.}, \au{Schoppa, W.} \& \au{Kim, J.}}
  \yr{1997}  \at{Coherent structures near the wall in a turbulent channel
  flow}.  \jt{J. Fluid Mech.}  \bvol{332},  \pg{185--214}.

\bibitem[Johansson {\em et~al.\/}(1991)Johansson, Alfredsson \&
  Kim]{johansson1991}
{\sc \au{Johansson, A.~V.}, \au{Alfredsson, P.~H.} \& \au{Kim, J.}} \yr{1991}
  \at{Evolution and dynamics of shear-layer structures in near-wall
  turbulence}.  \jt{J. Fluid Mech.}  \bvol{224},  \pg{579--599}.

\bibitem[Kevin {\em et~al.\/}(2019{\natexlab{{\em a\/}}})Kevin, Monty \&
  Hutchins]{kevin2019b}
{\sc \au{Kevin}, \au{Monty, J.} \& \au{Hutchins, N.}} \yr{2019{\natexlab{{\em
  a\/}}}}  \at{The meandering behaviour of large-scale structures in turbulent
  boundary layers}.  \jt{J. Fluid Mech.}  \bvol{865},  \pg{R1}.

\bibitem[Kevin {\em et~al.\/}(2019{\natexlab{{\em b\/}}})Kevin, Monty \&
  Hutchins]{kevin2019a}
{\sc \au{Kevin}, \au{Monty, J.} \& \au{Hutchins, N.}} \yr{2019{\natexlab{{\em
  b\/}}}}  \at{Turbulent structures in a statistically three-dimensional
  boundary layer}.  \jt{J. Fluid Mech.}  \bvol{859},  \pg{543--565}.

\bibitem[Kevin {\em et~al.\/}(2017)Kevin, Monty, Bai, Pathikonda, Nugroho,
  Barros, Christensen \& Hutchins]{kevin2017}
{\sc \au{Kevin}, \au{Monty, J.~P.}, \au{Bai, H.~L.}, \au{Pathikonda, G.},
  \au{Nugroho, B.}, \au{Barros, J.~M.}, \au{Christensen, K.~T.} \&
  \au{Hutchins, N.}} \yr{2017}  \at{Cross-stream stereoscopic particle image
  velocimetry of a modified turbulent boundary layer over directional surface
  pattern}.  \jt{J. Fluid Mech.}  \bvol{813},  \pg{412--435}.

\bibitem[Kim \& Adrian(1999)]{kim1999}
{\sc \au{Kim, K.~C.} \& \au{Adrian, R.~J.}} \yr{1999}  \at{Very large-scale
  motion in the outer layer}.  \jt{Phys. Fluids}  \bvol{11}~(2),
  \pg{417--422}.

\bibitem[Lee \& Sung(2011)]{lee2011}
{\sc \au{Lee, J.~H.} \& \au{Sung, H.~J.}} \yr{2011}  \at{Very-large-scale
  motions in a turbulent boundary layer}.  \jt{J. Fluid Mech.}  \bvol{673},
  \pg{80--120}.

\bibitem[Lee {\em et~al.\/}(2019)Lee, Sung \& Adrian]{leejh2019}
{\sc \au{Lee, J.~H.}, \au{Sung, H.~J.} \& \au{Adrian, R.~J.}} \yr{2019}
  \at{Space-time formation of very-large-scale motions in turbulent pipe flow}.
   \jt{J. Fluid Mech.}  \bvol{881},  \pg{1010--1047}.

\bibitem[{Lozano-Dur{\'a}n} {\em et~al.\/}(2012){Lozano-Dur{\'a}n}, Flores \&
  Jim{\'e}nez]{lozano-duran2012}
{\sc \au{{Lozano-Dur{\'a}n}, A.}, \au{Flores, O.} \& \au{Jim{\'e}nez, J.}}
  \yr{2012}  \at{The three-dimensional structure of momentum transfer in
  turbulent channels}.  \jt{J. Fluid Mech.}  \bvol{694},  \pg{100--130}.

\bibitem[Lumley(1967)]{lumley1967}
{\sc \au{Lumley, J.~L.}} \yr{1967} The structure of inhomogeneous turbulent
  flow.  \bt{In {\em Atmospheric Turbulence and Radio Wave Propagation\/} (ed.
  \ed{A.~M. Yaglom \& V.~I. Tatarski})},  \pg{pp. 166--178}.

\bibitem[Marusic {\em et~al.\/}(2010)Marusic, Mathis \& Hutchins]{marusic2010}
{\sc \au{Marusic, I.}, \au{Mathis, R.} \& \au{Hutchins, N.}} \yr{2010}
  \at{Predictive model for wall-bounded turbulent flow}.  \jt{Science}
  \bvol{329},  \pg{193--196}.

\bibitem[Medjnoun {\em et~al.\/}(2018)Medjnoun, Vanderwel \&
  Ganapathisubramani]{medjnoun2018}
{\sc \au{Medjnoun, T.}, \au{Vanderwel, C.} \& \au{Ganapathisubramani, B.}}
  \yr{2018}  \at{Characteristics of turbulent boundary layers over smooth
  surfaces with spanwise heterogeneity}.  \jt{J. Fluid Mech.}  \bvol{838},
  \pg{516--543}.

\bibitem[Meyer {\em et~al.\/}(2007)Meyer, Pedersen \& {\"O}zcan]{meyer2007}
{\sc \au{Meyer, K.~E.}, \au{Pedersen, J.~M.} \& \au{{\"O}zcan, O.}} \yr{2007}
  \at{A turbulent jet in crossflow analysed with proper orthogonal
  decomposition}.  \jt{J. Fluid Mech.}  \bvol{583},  \pg{199–227}.

\bibitem[Modesti {\em et~al.\/}(2018)Modesti, Pirozzoli, Orlandi \&
  Grasso]{modesti2018}
{\sc \au{Modesti, D.}, \au{Pirozzoli, S.}, \au{Orlandi, P.} \& \au{Grasso, F.}}
  \yr{2018}  \at{On the role of secondary motions in turbulent square duct
  flow}.  \jt{J. Fluid Mech.}  \bvol{847},  \pg{R1}.

\bibitem[Nikora {\em et~al.\/}(2019)Nikora, Stoesser, Cameron, Stewart,
  Papadopoulos, Ouro, McSherry, Zampiron, Marusic \& Falconer]{nikora2019}
{\sc \au{Nikora, V.~I.}, \au{Stoesser, T.}, \au{Cameron, S.~M.}, \au{Stewart,
  M.}, \au{Papadopoulos, K.}, \au{Ouro, P.}, \au{McSherry, R.}, \au{Zampiron,
  A.}, \au{Marusic, I.} \& \au{Falconer, R.~A.}} \yr{2019}  \at{Friction factor
  decomposition for rough-wall flows: theoretical background and application to
  open-channel flows}.  \jt{J. Fluid Mech.}  \bvol{872},  \pg{626--664}.

\bibitem[Nugroho {\em et~al.\/}(2013)Nugroho, Hutchins \& Monty]{nugroho2013}
{\sc \au{Nugroho, B.}, \au{Hutchins, N.} \& \au{Monty, J.~P.}} \yr{2013}
  \at{Large-scale spanwise periodicity in a turbulent boundary layer induced by
  highly ordered and directional surface roughness}.  \jt{Int. J. Heat Fluid
  Flow}  \bvol{41},  \pg{90--102}.

\bibitem[Raupach \& Shaw(1982)]{raupach1982}
{\sc \au{Raupach, M.~R.} \& \au{Shaw, R.~H.}} \yr{1982}  \at{Averaging
  procedure for flow within vegetation canopies}.  \jt{Bound.-Layer Meteorol.}
  \bvol{22},  \pg{79--90}.

\bibitem[Schoppa \& Hussain(2002)]{schoppa2002}
{\sc \au{Schoppa, W.} \& \au{Hussain, F.}} \yr{2002}  \at{Coherent structure
  generation in near-wall turbulence}.  \jt{J. Fluid Mech.}  \bvol{453},
  \pg{57--108}.

\bibitem[Sillero {\em et~al.\/}(2014)Sillero, Jim\'{e}nez \&
  Moser]{sillero2014}
{\sc \au{Sillero, J.~A.}, \au{Jim\'{e}nez, J.} \& \au{Moser, R.~D.}} \yr{2014}
  \at{Two-point statistics for turbulent boundary layers and channels at
  \uppercase{R}eynolds numbers up to $\delta^{+} \approx 2000$}.  \jt{Phys.
  Fluid}  \bvol{26},  \pg{105109}.

\bibitem[de~Silva {\em et~al.\/}(2018)de~Silva, Kevin, Baidya, Hutchins \&
  Marusic]{desilva2018}
{\sc \au{de~Silva, C.~M.}, \au{Kevin}, \au{Baidya, R.}, \au{Hutchins, N.} \&
  \au{Marusic, I.}} \yr{2018}  \at{Large coherence of spanwise velocity in
  turbulent boundary layers}.  \jt{J. Fluid Mech.}  \bvol{847},  \pg{161--185}.

\bibitem[Sirovich(1987)]{sirovich1987}
{\sc \au{Sirovich, L.}} \yr{1987}  \at{Turbulence and the dynamics of coherent
  structures. \uppercase{P}art i: \uppercase{C}oherent structures}.  \jt{Q.
  Appl. Maths.}  \bvol{45}~(3),  \pg{561--571}.

\bibitem[Tomkins \& Adrian(2003)]{tomkins2003}
{\sc \au{Tomkins, C.~D.} \& \au{Adrian, R.~J.}} \yr{2003}  \at{Spanwise
  structure and scale growth in turbulent boundary layers}.  \jt{J. Fluid
  Mech.}  \bvol{490},  \pg{37--74}.

\bibitem[Townsend(1976)]{townsend1976}
{\sc \au{Townsend, A.~A.}} \yr{1976} {\em The Structure of Turbulent Shear
  Flow\/}, 2nd edn.  \publ{Cambridge: Cambrige University Press}.

\bibitem[Vanderwel {\em et~al.\/}(2019)Vanderwel, Stroh, Kriegseis, Frohnapfel
  \& Ganapathisubramani]{vanderwel2019}
{\sc \au{Vanderwel, C.}, \au{Stroh, A.}, \au{Kriegseis, J.}, \au{Frohnapfel,
  B.} \& \au{Ganapathisubramani, B.}} \yr{2019}  \at{The instantaneous
  structure of secondary flows in turbulent boundary layers}.  \jt{J. Fluid
  Mech.}  \bvol{865},  \pg{845--870}.

\bibitem[Waleffe(2001)]{waleffe2001}
{\sc \au{Waleffe, F.}} \yr{2001}  \at{Exact coherent structures in channel
  flow}.  \jt{J. Fluid Mech.}  \bvol{435},  \pg{93--102}.

\bibitem[Wangsawijaya(2020)]{wangsawijaya2020-PHD}
{\sc \au{Wangsawijaya, D.~D.}} \yr{2020}  \at{Time-varying secondary flows in
  turbulent boundary layers over surfaces with spanwise heterogeneity}. PhD
  thesis, University of Melbourne.

\bibitem[Wangsawijaya {\em et~al.\/}(2020)Wangsawijaya, Baidya, Chung, Marusic
  \& Hutchins]{wangsawijaya2020}
{\sc \au{Wangsawijaya, D.~D.}, \au{Baidya, R.}, \au{Chung, D.}, \au{Marusic,
  I.} \& \au{Hutchins, N.}} \yr{2020}  \at{The effect of spanwise wavelength of
  surface heterogeneity on turbulent secondary flows}.  \jt{J. Fluid Mech.}
  \bvol{894},  \pg{A7}.

\bibitem[Wu {\em et~al.\/}(2012)Wu, Baltzer \& Adrian]{wu2012}
{\sc \au{Wu, X.}, \au{Baltzer, J.~R.} \& \au{Adrian, R.~J.}} \yr{2012}
  \at{Direct numerical simulation of a 30\uppercase{$R$} long turbulent pipe
  flow at \uppercase{$R$}$^+ = 685$: large- and very large-scale motions}.
  \jt{J. Fluid Mech.}  \bvol{698},  \pg{235--281}.

\bibitem[Zampiron {\em et~al.\/}(2020)Zampiron, Cameron \&
  Nikora]{zampiron2020}
{\sc \au{Zampiron, A.}, \au{Cameron, S.} \& \au{Nikora, V.}} \yr{2020}
  \at{Secondary currents and very-large-scale motions in open-channel flow over
  streamwise ridges}.  \jt{J. Fluid Mech.}  \bvol{887},  \pg{A17}.

\end{thebibliography}


\end{document}